\documentclass[format=acmsmall, manuscript, screen]{acmart}

\AtBeginDocument{%
  \providecommand\BibTeX{{%
    \normalfont B\kern-0.5em{\scshape i\kern-0.25em b}\kern-0.8em\TeX}}}
    
\usepackage[utf8]{inputenc}
\usepackage[colorinlistoftodos]{todonotes}
\usepackage{amsmath}
\usepackage{amssymb}
\usepackage{geometry}
\usepackage[colorinlistoftodos]{todonotes}
\usepackage[toc,page]{appendix}
\usepackage{tikz}
\usepackage{collcell}
\usepackage{soul}

\usepackage{xcolor}
\usepackage{array}
\usepackage{bm}
\usepackage{float}
\usepackage{graphicx,grffile}
\usepackage{subfigure}
\usepackage{verbatim}
\usepackage[noabbrev,capitalize]{cleveref}
\usepackage{wrapfig}

\setstcolor{red}
\title{BISTRO: Berkeley Integrated System for Transportation Optimization}

\author{Sidney A.~Feygin}
\authornote{At the time of writing, Lee and Gupta were full-time employees at Uber. Feygin, Lazarus, Forscher, Golfier-Vetterli, and Bayen contributed to this work partially as Uber contractors.}
\email{sid.feygin@gmail.com}
\affiliation{
    \institution{Uber Technologies, Inc.}
}

\author{Jessica R.~Lazarus}
\authornotemark[1]
\email{jlaz@berkeley.edu}
\author{Edward H.~Forscher}
\authornotemark[1]
\email{forscher@berkeley.edu}
\affiliation{
    \institution{University of California, Berkeley; Uber Technologies, Inc.}
}

\author{Valentine Golfier-Vetterli}
\authornotemark[1]
\email{vgolfi@ext.uber.com}
\author{Jonathan W. Lee}
\authornotemark[1]
\email{jonny@uber.com}
\author{Abhishek Gupta}
\authornotemark[1]
\email{agupta@uber.com}
\affiliation{
    \institution{Uber Technologies, Inc.}
}

\author{Rashid A.~Waraich}
\email{waraich.rashid@gmail.com}
\author{Colin J.R.~Sheppard}
\email{colin.sheppard@lbl.gov}
\affiliation{
    \institution{Lawrence Berkeley National Laboratory}
}

\author{Alexandre M.~Bayen}
\authornotemark[1]
\affiliation{
    \institution{University of California, Berkeley; Uber Technologies, Inc.; Lawrence Berkeley National Laboratory}
}
\email{bayen@berkeley.edu}

\renewcommand{\shortauthors}{Feygin, \textit{et al.}}

\date{\today}

\begin{document}

% \linenumbers

\begin{abstract}
The current trend towards urbanization and adoption of flexible and innovative mobility technologies will have complex and difficult-to-predict effects on urban transportation systems. Comprehensive methodological frameworks that account for the increasingly uncertain future state of the urban mobility landscape do not yet exist. Furthermore, few approaches have enabled the massive ingestion of urban data in planning tools capable of offering the flexibility of scenario-based design.

This article introduces BISTRO, a new open source transportation planning decision support system that uses an agent-based simulation and optimization approach to anticipate and develop adaptive plans for possible technological disruptions and growth scenarios. The new framework was evaluated in the context of a machine learning competition hosted within Uber Technologies, Inc., in which over 400 engineers and data scientists participated. For the purposes of this competition, a benchmark model, based on the city of Sioux Falls, South Dakota, was adapted to the BISTRO framework. An important finding of this study was that in spite of rigorous analysis and testing done prior to the competition, the two top-scoring teams discovered an unbounded region of the search space, rendering the solutions largely uninterpretable for the purposes of decision-support. On the other hand, a follow-on study aimed to fix the objective function. It served to demonstrate BISTRO's utility as a human-in-the-loop cyberphysical system: one that uses scenario-based optimization algorithms as a feedback mechanism to assist urban planners with iteratively refining objective function and constraints specification on intervention strategies. The portfolio of transportation intervention strategy alternatives eventually chosen achieves high-level regional planning goals developed through participatory stakeholder engagement practices.
\end{abstract}

% \ccsdesc[300]{Computing methodologies~Systems theory}
% \ccsdesc[300]{Computing methodologies~Visual analytics}
% \ccsdesc[300]{Computing methodologies~Artificial life}
% \ccsdesc[300]{Computing methodologies~Simulation environments}
% \ccsdesc[100]{Computing methodologies~Network science}
% \ccsdesc[100]{Computing methodologies~Multiscale systems}
% \ccsdesc[100]{Computing methodologies~Scientific visualization}
% \ccsdesc[100]{Computing methodologies~Simulation evaluation}
% \ccsdesc[500]{Applied computing~E-government}
% \ccsdesc[300]{Applied computing~Computer-aided design}
% \ccsdesc[300]{Applied computing~Multi-criterion optimization and decision-making}
% \ccsdesc[100]{Applied computing~Cartography}
% \ccsdesc[100]{General and reference~Experimentation}
% \ccsdesc[100]{Security and privacy~Privacy protections}


% Page budget: 
% \begin{enumerate}
%     \item Intro and problem statement: 2-3 pages
%     \item Background: 2 pages [(1) Transportation planning and policy: (0.5 page), (2) ABM: general science (0.5 page), Matsim: (0.25 page), BEAM (0.75 page) overview, architecture and motivation.]
%     \item BISTRO: 4-5 pages [(1) architecture (1 page), (2) scoring (0.5 page with ample references to the other doc: general process for designing scoring functions; just the essential KPIs and reference to the doc for Sioux Falls); (3) change current 3.3 into a true input section (0.5 ish pages); (4) Output analysis; what are the signicant statistics out of BEAM (Aggregation of agent behavior informing algorithm design), using domain knowledge (0.5 page); (5) Visualization - Jupyter work, functionalities (1 page); (6) Software architecture: (0.5 - 1 page). 
%     \item Case Study: Crowdsourced Planning: 4-6 pages [(1) Sioux Faux: (1 page), (2) Contest participation: (0.5 page), Results: (2-4 pages)]; 
%     \item Ongoing/Future Work (1 page): merge into the previous section: call ``reproducible research'' $\rightarrow$ emphasize (a) original (b) fix (just cost function fix), (c) new BAUs (incl. new demographics and KPIs)
%     \item San Francisco (1-2 pages max). 
%     \item Conclusion: (10 lines)
% \end{enumerate}

\maketitle

\section{Introduction} 
As modern transportation systems undergo a period of intense technological change, researchers, practitioners, and policymakers are seeking to understand how long-term trends towards vehicle digitalization, automation, electrification, as well as the emerging sharing economy will shape the future day-to-day dynamics of human mobility in cities worldwide.

% shaping the operational, social, financial, and environmental dynamics of human mobility in cities worldwide. 

Likewise, rapid advances in computing as well as the advent of metropolitan scale data mining and pattern recognition (\textit{i.e.}, machine learning) have, for the first time, made it possible to characterize urban traffic flows based on movement traces from millions of individual travelers. These methods fuse passively collected spatiotemporal trajectories derived from smartphone data with static census data in order to map travel patterns to sociodemographic characteristics. Specialized traffic models can thereby be trained using feature-rich representations of human mobility.

While such predictive models have proven effective in nearcasting congestion events  \cite{yin2018activity}, purely data-driven methods often fail to generalize when applied to the task of forecasting the effect of transportation policy strategies on future demand--particularly when anticipated changes in urban sociodemographics, regulatory policy, and mobility technologies as well as interactions between these layers cannot themselves be predicted with a high degree of certainty \cite{Zheng2013}. In other words, the covariates of statistical models may not capture the numerous interacting physical and behavioral components (\textit{i.e.}, \textit{network effects}) influencing transportation supply and demand.

\textit{Agent-based modeling and simulation} (ABMS) techniques and software, on the other hand, are beginning to be used as flexible decision-support tools by planning groups, technologists, and regulatory agencies to facilitate forecasting the short- and long-term implications of transportation system interventions. That is, by simulating models of traveler decision-making in the context of a landscape of future plausible ``states of the world'', stakeholders can better resolve uncertainty over how transportation system interventions may fare. For example, ABMS enables the investigation of the extent to which citizens may adopt on-demand autonomous vehicles, as well as how to sustainable fleet operation \cite{Hornia,Horl2016}.

% In order to map aggregate model outputs to corresponding metrics derived from real-world observations, researchers are increasingly exploring calibration and validation protocols that make use of so-called ``big data''-driven methods, such as the statistical learning techniques indicated above \cite{anda2017transport}.

Even with these advances, selecting interventions that balance competing transportation system policy objectives remains a difficult and contentious process. Current methods to identify the best alternative under a given scenario often involves the use of Monte Carlo methods to evaluate different policy options. Such approaches can be extremely computationally expensive for large ABMS implementations and are unlikely to yield an optimal solution. Moreover, transportation researchers and practitioners often develop scenario-specific and/or geography-specific simulation models with limited integration of the efficient and highly generalizable methods developed in the \textit{artificial intelligence} (AI) and \textit{machine learning} (ML) communities. 
 
The \textit{Berkeley Integrated System for TRansportation Optimization} (BISTRO) is an open-source framework and software toolkit designed to address the increasingly complex problems arising in transportation systems worldwide. BISTRO does this through an innovative human-machine collaboration approach, using state-of-the-art optimization algorithms to efficiently automate the search for policy interventions that achieve good performance over diverse transportation system scenarios and stakeholder objectives. BISTRO includes an ABMS system and scenario development pipeline to build empirically-calibrated simulations of travel demand in metropolitan transportation systems. After a calibrated scenario has been developed, BISTRO enables use of state-of-the-art optimization algorithms to identify system interventions that best align with policy and planning objectives. Users can deploy BISTRO to enable distributed development of algorithms that rapidly optimize a feasible set of policy and investment decisions. Once one or more desirable solutions are found, BISTRO provides a suite of analysis and visualization tools to empower citizens, transportation system planners and engineers, private entities, and governments to better understand and collaborate on developing strategies that achieve equitable access to and sustainable use of current and emerging mobility services.

The rest of this article is organized as follows: \cref{sec:background} provides a brief overview of the current state of transportation planning, ABMS tools, and transportation optimization tools, respectively;  \cref{sec:architecture} presents BISTRO, covering the system architecture, scoring function design, inputs, outputs, analysis capabilities, and performance characteristics; \cref{sec:pilot_study} details an initial pilot study, updates to BISTRO based upon the pilot, and a few algorithmic solution approaches; and \cref{sec:conclusion} offers a short conclusion.

\section{Background}
\label{sec:background}
\subsection{Transportation Planning and Policy Decision Support Systems}
\label{subsec:trans_planning_background}

In the United States, the primary outcome of efforts to design, model, and communicate the impacts of proposed policy interventions and infrastructure investments on transportation systems are Metropolitan \textit{Regional Transportation Plans} (RTPs) and State \textit{Long-Range Transportation Plans} (LRTPs). These plans are forward-looking, long-term (20+ year time horizons) and have been a federally mandated task since the Federal-Aid Highway Act of 1962. Recently, state departments of transportation and \textit{Metropolitan Planning Organizations} (MPOs)---the entities tasked with producing RTPs every four-to-five years---have been shifting towards performance-based planning and programming (PBPP) frameworks \cite{grant2013performance}.

 As part of RTP development using PBPP guidelines, an MPO or related agency will typically conduct a community engagement process to identify one or more \textit{visions} or \textit{goals} shared by stakeholders and the public describing a desired future state of the regional transportation system (e.g., safe roadways, accessible transit, environmental stewardship) \cite{DOT_planning, DOT_public_engagement, grant2013performance}. Measurable \textit{objectives} are defined together with quantitative \textit{performance indicators} in order to evaluate the extent to which alternative \textit{strategies} consisting of policy interventions or infrastructure investments could make progress towards achievement of a goal \cite{DOT_planning, grant2013performance}. Typically, the impact of these projects on \textit{key performance indicators} (KPIs) of transportation system performance will be forecast using an analytic, data-driven model of travel demand\footnote{Traditionally, the travel demand modeling process consists of four main steps: 1) trip generation to and from all analysis zones, 2) trip distribution (or matching origins and destinations, often using a gravity model), 3) assigning  traveler mode choice based upon individual preferences and alternative characteristics, and 4) route assignment, of trips onto physical network links; this is referred to as the four-step model. Many MPOs and other agencies are moving towards disaggregate, activity-based, or person-centric models of daily activity, rather than aggregate approaches operating on the zonal level \cite{DOT_planning}.}.

The domain expertise necessary to understand the functionality of models of transportation demand may result in recommendations that are frustratingly opaque to public interpretation \cite{DOT_public_engagement}.  Explanations of the inner workings of the modeling process are not presented during public collaboration meetings and questions regarding the validity of models have led to lawsuits over the lack of publicly available information on model specifics \cite{Sciara2017}. Often, contractual obligations or proprietary data formats used by the consulting firms tasked with developing planning software further restrict public access. The inability or unwillingness on the part of MPOs to promote flexible open-source software licensing strategies discourages independent investigations (by, for example, academic or public interest groups) to verify that the theory, equations, algorithms, and data comprising a model match its software implementation. This lack of operational transparency not only subverts accountability, but also impedes the rapid transfer of innovative modeling methods and technologies across and even within agencies  \cite{transpo_planning_process, GAO_MPO}. 

To address criticism concerning model transparency and explainability, recent transportation and land-use planning organizations have begun to pilot software that visualizes the impacts of alternatives and provide fora for public input via collaborative simulation platforms \cite{DOT_planning, auld2016polaris, Ocalir-Akunal2016,Khan2014,azevedo2018tripod}. An interesting example is the UrbanAPI project, which includes a web-based 3-dimensional virtual reality visualization of the impacts of urban growth scenarios at the scale of individual neighborhoods \cite{Khan2014}. While initial efforts have focused on the usability requirements of these tools, their broader impact has been limited due to narrowly defined geographic contexts or specialization for regional system objectives \cite{Postorino}. 

One project that does have broad support across several MPOs is ActivitySim, which currently being developed by the Association of Metropolitan Planning Organizations as an open source activity-based travel modeling platform \cite{activitysim}. ActivitySim may be used in concert with synthetic households, synthetic persons,  employment data, land use data, and network performance data  (e.g., travel times by mode and time of day, costs, and transfers) to generate a full-scale model for a city's travel demand. ActivitySim may be seen as a complementary project to our own in that ActivitySim model outputs can be used as input data for a BISTRO scenario.

\subsection{ABMS of Transportation Systems}
\label{subsec:ABMSTrans}
% This subsection answers the following questions: 
% \begin{enumerate}
%     \item What is agent-based modeling and simulation (ABMS)?
%     \item How is ABMS used in a transportation planning setting and what are some examples?
%     \item What are the benefits and possible criticisms of simulating transportation decision-making using ABMS?
% \end{enumerate}

% \subsubsection{Preliminaries on ABMS}

% Intro to ABMS with focus on their use to study complex adaptive systems (CAS). 
Developing models that replicate how urban systems operate and evolve has been a major focus of transportation engineers, urban planners, and geographers. Trip-based methods, such as the traditional four-step model used by MPOs, are specified at aggregate geographic or categorical levels rather than at the level of individual decision-makers. This level of aggregation can limit the ability of such models to explain complex individual decisions.
%impairs their ability to forecast how future and emerging transportation technology, incentives, and policies will shape traffic behavior. 
Some MPOs have begun to adopt a more behaviorally-descriptive activity-based approach \cite{Castiglione2014}. In contrast to trip-based models, activity-based methods represent more comprehensive links between activity scheduling, mode choice, social interaction, and spatiotemporal constraints \cite{Castiglione2014}. Agent-based models and simulations (ABMS) of transportation demand are capable of replicating observed \textit{macroscopic} traffic patterns by simulating the \textit{microscopic} decision-making behavior of a synthetic population of software agents as they execute their daily travel plans on a virtual model of the transportation system. When \textit{calibrated} to ground-truth data by modifying only global parameters characterizing the embedded choice model, agent-based frameworks represent parsimonious descriptions of regional travel demand. Consequently, ABMS are capable of accurately capturing shifts in macro patterns when forecast changes in transportation infrastructure, policies, demographics, and vehicle ownership are introduced to the virtual travel environment \citep{Zheng2013}. 

In the past three decades, several multi-agent frameworks such as TRANSIMS \cite{smith1995transims}, MATSim \cite{Hornia}, SUMO \cite{behrisch2011sumo}, and POLARIS \cite{auld2016polaris} have been developed and widely adapted for numerous applications in transportation and land use planning and research \cite{Castagnari,Kamel2019,Zheng2013,Postorino}. MPOs often couple urban development simulation models\footnote{Examples of these include the \textit{Integrated Transportation, Land Use, and Environment} model (ILUTE, \cite{salvini2005ilute}) and UrbanSim \cite{waddell2002urbansim}.} with microscopic agent-based transport models to better understand how predicted changes in population growth, land-use, real-estate development, and resource markets will co-evolve with changes in the transportation system \cite{nicolai2012coupling,Moeckel2018}. ABMS of transportation systems can take different population configuration files as inputs, giving planners and modelers the ability to simulate how long-term urbanization processes can be shaped by the daily transportation decisions of individuals (possibly in the presence of alternative policy interventions and new mobility technologies) \cite{nicolai2010coupling,Moeckel2018}. Increasingly, these open-source platforms are enabling use of publicly available data to create transparent and replicable input preparation pipelines, reducing the cost and effort of a variety of urban planning tasks \cite{ziemke2016towards}. Since BISTRO primarily relies on BEAM and BEAM itself incorporates many aspects of MATSim, the remainder of this section takes a closer look at these two software frameworks' purpose, functionality, and computational characteristics.

\begin{wrapfigure}{R}{0.5\textwidth}
\includegraphics[width=0.5\textwidth]{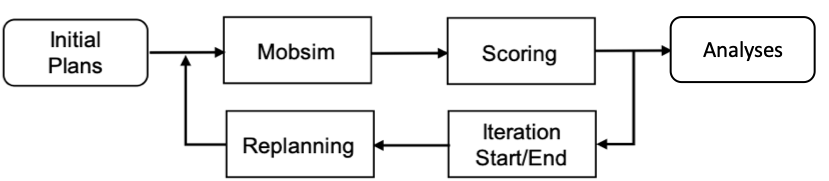}
\caption{Conceptual process of MATSim. It iteratively evaluates and mutates a proportion of agent plans until the utility of plans no longer improves. At this point, the system is said to have reached a \textit{stochastic user equilibrium}. For further details, see \cite{Hornia}.}
\label{fig:matsim_workflow}
\end{wrapfigure}

\paragraph{MATSim}  MATSim is an ABMS framework developed by teams at ETH Z{\"u}rich and TU Berlin \cite{Hornia}. \hbox{MATSim} enables simulating the travel behavior of millions of individual agents, representing a synthetic population of urban travelers. At the heart of MATSim is a co-evolutionary algorithm that iteratively executes, evaluates, and mutates (i.e., replans) the \textit{daily} activity schedules of agents (see \Cref{fig:matsim_workflow}). The end result of this process is an equilibrium between network supply and travel demand--resulting in realistic congestion patterns as agents compete for limited space on a virtual road network.

Besides road network and transit data, the key input to MATSim is the agent \textit{population}. This file encodes a set of unrealized plans (one for each agent) consisting of the start times, types (e.g., ``Home'', ``Work'', ``Shopping'', ``School'', etc.) and locations of various significant activities\footnote{The population file is often generated from the output of an activity-based travel demand model.}. A mobility simulation (\texttt{MobSim}) executes these plans on a virtual road network. While, at first, agents only drive and/or walk to activities, additional modes may be introduced during plan mutation (explained below). Each agent's plan is then scored\footnote{This score can be interpreted as econometric utility. It is measured by a linear model that assigns negative value to time spent traveling and positive value to time spent at activities.}. Copies of evaluated plans are stored in a limited-size array, representing the agent's memory. At the start of the subsequent iteration, a portion of agents in the population are chosen to have a randomly selected mutation strategy applied to a plan drawn from each of their memories. Examples of plan mutations include changing activity start time, mode of travel for a tour (or subtour), and selecting a different route based on previously experienced link travel times. Optional extensions to MATSim model further behavioral dimensions such as parking choice, group travel, and vehicle sharing.

The algorithm converges once agents are no longer able to improve the score of plans in their memory, at which point MATSim produces a series of statistics and outputs describing the aggregate performance of system components as well as a snapshot of all events that occurred over the course of the simulation. Events can be processed to derive the actual paths and travel times realized by each agent and each vehicle, as well as other data reflecting simulation performance.

% \textit{Why was BEAM created and how does it innovate over MATSim architecture? What are salient features in BEAM that make it particularly useful for analysis of innovative long and short term transportation policies?}
\paragraph{BEAM} The \textit{Behavior Energy Autonomy and Mobility} (BEAM) framework is a multi-agent travel demand simulation framework developed at \textit{Lawrence Berkeley National Laboratory} (LBNL) \cite{us_department_of_energy_energy_2018}. While the overall learning and traffic assignment mechanism is similar to MATSim's co-evolutionary algorithm, added functionality in BEAM is specifically focused on helping users understand the impacts of new and emerging travel modes on limited capacity resource markets. This subsection describes essential features of BEAM that led to its selection as the core simulation engine of BISTRO.

% Due to the expanding variety of mobility alternatives, multiple factors driving down the propensity to drive alone (such as environmental awareness, gas prices, and availability of novel and more convenient or less expensive alternatives), as well as real-time information provided via smartphone apps, travelers are increasingly combining multiple travel modes into one trip. 

BEAM is closely integrated with the transit service capabilities of the R5 routing engine, which include \textit{General Transit Feed Specification} (GTFS) file processing and routing based on multiobjective variations of the RAPTOR algorithm \cite{Delling2008}. Transit may be combined with other modes modeled in BEAM such as autonomous vehicles, on-demand rides, e-bikes, and scooters, enabling agents to make realistic, multimodal mobility decisions. In order to provide agents with information about the time and monetary costs of different travel options, R5 computes the lowest generalized cost path (based on travel time estimates from the mobility simulation) for the corresponding mode(s) available to the agent for the trip.\footnote{For detailed information about the R5 router, see \cite{conway2017evidence}.} 

Unlike the replanning mechanism in MATSim, which only mutates plans \textit{between} consecutive iterations, agents in BEAM are designed to adapt to changing conditions \textit{during} an iteration according to what is known as a \textit{within-day} or \textit{online} model\footnote{While MATSim has a within-day mode, much of the functionality enabled by extension modules representing emerging mobility technologies assumes that replanning happens between iterations.}. Thus agents can make unplanned and time-sensitive choices about how to maximize the score of their travel plans while competing for limited resources that vary in availability over time. For example, an agent that chooses a transit mode may be denied access to an overfull bus, requiring the agent to make a mid-trip change to their itinerary. The agent could then choose to wait for the next bus or hail a ride from the point of departure (if a driver is available nearby). The BEAM software architecture addresses the performance and complexity challenges of integrating new models of within-day dynamics by implementing agents as \textit{actors}, as defined within the  \textit{actor-based model of concurrency}\footnote{Like objects in the object-oriented programming paradigm, actors encapsulate state and behavior. However, unlike the object model, actors do not share computer memory. Instead, each actor encapsulates its own thread of execution and interacts with one other actors using messages. An actor may send message to other actors without blocking. Each actor processes messages synchronously in the order received; however, computation is scheduled asynchronously over multiple actors. Thus, the actor-based model of computation obviates the need for locking mechanisms commonly used to synchronize state among interdependent objects. Consequently, reasoning about agent behavior using actors can allow researcher developers to focus on implementing novel models and applications rather than debugging threads and locks \cite{agha2001actors}}. 

Like MATSim, BEAM enables users to model realistic variations in travel preferences predicated on agent characteristics (which are themselves derived from sociodemographic statistics computed on census data and travel surveys). However, the mechanism for selection of travel alternatives differs significantly from MATSim's in order to model within-day decision-making. Specifically, the probability of selecting an alternative (route, mode, parking choice, and refueling decision) is represented in BEAM according to a multinomial logit model \cite{Train2003,ben1985discrete}. That is, among several distinct travel schedules, agents are exponentially more likely to select the option that maximizes their enjoyment of important activities while reducing time and money spent traveling between activity locations. The choices made by BEAM agents and their corresponding scores comprise a BEAM plan, which is stored in memory following its execution. Thus, in the transit choice scenario described above, an agent’s response to a full bus could incorporate multiple downstream choices within one iteration rather than wait for a score penalty to propagate an optimal response through the selected plan (i.e., using plan mutations that take place over the course of multiple iterations)\footnote{As in MATSim, the highest scoring BEAM plans are more likely to be re-evaluated and possibly selected for mutation. However, in contrast to MATSim, plans selected for mutation are cleared of all choices and re-evaluated within the context of the current iteration's transient state. Empirically, we find that this approach reduces the number of iterations needed to reach equilibrium.}.

\subsection{Simulation-based Optimization of Transportation Systems}
\label{subsec:simulation-problem}

\subsubsection{Optimization-based formulation of the planning problem} The problem class solved by the BISTRO framework can be characterized as simulation-based optimization of large urban transportation systems. It can be symbolically formulated as an optimization problem:
\begin{equation}
\label{eq:opt_problem_objective}
\underset{\vec{d}\in\mathcal{D}}{\textnormal{minimize}} \ f(\vec{d},\vec{x};\vec{z})\equiv\mathbb{E}[F(\vec{d},\vec{x};\vec{z})]
\end{equation}
constrained by simulation outcomes and design constraints, \textit{i.e.},
\begin{equation}
\label{eq:constraints}
    \begin{cases}
        \vec{x}=B(\vec{d};\vec{z}),\\
        g(\vec{d};\vec{z})=0,
    \end{cases}
\end{equation}
respectively, where the objective, $f$, is defined as the expected value of a stochastic performance measurement function, $F$. The deterministic decision vector, $\vec{d}$, is chosen from a search space $\mathcal{D}$, which may be continuous, categorical, combinatorial, or conditional. In the BISTRO context, the decision variables, $\vec{d}$, are the user-defined inputs that control policy levers within the transportation system. For example, $\vec{d}$ may specify the fare or vehicle types for specific public transit routes. The exogenous variables, $\vec{z}$, are the configuration inputs that determine the parameters of the population synthesis, the parameters of the transportation network, and the parameters governing supply of transportation services. The endogenous variables, $\vec{x}$, are the outcomes of the simulation run using $\vec{d}$ and $\vec{z}$ as input. The vector $\vec{x}$ contains the details of agent and vehicle movements throughout the simulation run, such as mode choices, travel times, travel costs, and vehicle path traversals, that were realized during the simulation run, \textit{i.e.}, $\vec{x} = B(\vec{d}; \vec{z})$, where $B$ represents the BEAM simulator. It is assumed that the iterative simulation process described in \Cref{subsec:ABMSTrans} has achieved stationarity.\footnote{Individual optimization algorithms may relax this constraint in order to reduce compute time while potentially trading off reduced accuracy or increased stochasticity of simulation output statistics.} 

An important goal of BISTRO is to define objective functions that guide algorithms towards a range of solutions that represent \textit{interpretable} and implementable policy decisions. A critical safeguard against unrealistic outcomes is implemented in BISTRO by translating business rules about inputs into mathematical constraint functions, $g$, parameterized by the decision vector, $\hat{d}$. Finally, $F$, is computed as a convex combination of the score components that guide solutions towards the \textit{system objective} (as defined in \Cref{subsec:trans_planning_background}).\footnote{Due to variable amounts of nondeterminism and stochasticity inherent in ABMS, given fixed $\vec{d}$ and $\vec{z}$, the distribution of $f$ can be approximated using $n$ realizations of $F$ as $\ensuremath{\hat{f}\left(\vec{d},\vec{x};\vec{z}\right)=\frac{1}{n}\sum_{i=1}^{n}}F_{i}\left(\vec{d}_i,\vec{x}_i;\vec{z}\right)$. In practice, optimization usually proceeds with $n=1$ in order to identify promising (i.e., close to optimal) subsets of $\mathcal{D}$; however, when reporting final scores, one must carefully select $n$ such that variability in output values is adequately captured.} The score components are evaluated from key performance indicators (KPIs) of the system performance, which are calculated using the simulation outputs, and relevant inputs. Following \cite{osorio2013simulation}, we approximate the objective as the sample average of $r$ independent realizations of $F$:

\begin{equation}
    \label{eq:sample_average}
    \hat{f}(\vec{d},\vec{x};\hat{z})=\frac{1}{r}\sum_{i=1}^{r}F_i(\vec{d},\vec{x};\hat{z})
\end{equation}

The state space dynamics that govern the simulation-based optimization of an urban transportation system are highly complex and nonlinear, potentially containing several local minima. The lack of closed-form solutions to this class of optimization problem together with the computational expense associated with evaluating a single decision point make this class of problems particularly difficult to solve. In the next subsection, we describe several possible approaches to address this complexity.
 
% The movement of vehicles is determined by the physical dynamics of each network that define the effects of congestion on travel times, which in turn affect the supply available to travelers and thus impact the distribution of demand across the entire system. 

\subsubsection{Optimizing complex simulated systems: challenges and approaches} 

As described in \Cref{subsec:ABMSTrans}, agent-based micro- or meso-scale simulations of transportation systems model the interdependent choices of rational individuals as they navigate virtual representations of physical and human geographies. Calibrating such models to high-resolution GPS traces and other sensor data embedded in infrastructure makes them highly suitable for evaluating the outcomes of location-specific policy alternatives. The trade-offs in accounting for the heterogeneous preferences of millions of agents, are that 1) the simulation model is expensive to evaluate for different settings of $\vec{d}$, and 2) the complex relationship between network dynamics and agent behavior lead to stochastic, non-convex specifications of performance measure, $F$. Consequently, the efficient gradient-based methods used to optimize closed-form relaxations of mobility dynamics as well as data-driven models derived from historical movement patterns do not apply~\cite{osorio2013simulation}. Instead, generalized \textit{stochastic optimization} (SO) algorithms treat the simulator as a black box. Commonly used derivative-free SO approaches include grid search, random search, ranking and selection, metaheuristic, and \textit{metamodeling} techniques~\cite{conn2009introduction,Barton2006}. 

Metamodeling algorithms encompass a broad class of simulation-based optimization approaches. These approximate $F$ using a \textit{surrogate model}, $\mathcal{Q}$ that is less costly to evaluate. Flexible and computationally tractable representations such as polynomial splines are able to approximate any objective function; however, many simulation runs are still required to accurately fit the response surface of the underlying system~\cite{conn2009introduction,Barton2006}. 

\textit{Sequential model-based optimization} (SMBO) is a general metamodeling formalism that, given a history of previous evaluations,  $\mathcal{H}=\left\{\left(\vec{d_1},y_1\right),\ldots,\left(\vec{d_i},y_i\right)\right\}$, of observations $y_i = F\left(\vec{d}_i,\vec{x}_i;\vec{z}\right)$ at sample points in $\mathcal{D}$, selects the optimal next point $\vec{d}_{i+1}$ based on an approximation of $F$. To initialize SMBO, a small set of samples, $\left\{\vec{d}_1,\ldots,\vec{d}_i\right\}$ from $\mathcal{D}$ are selected using various experimental design techniques (e.g., random or Latin hypercube sampling). For each $\vec{d_i}$, evaluations of the expensive objective function, $F$ form an observation, which, together with $\vec{d}_i$ are appended to a historical dataset $\mathcal{H}$. Once $\mathcal{H}$ is initialized, SMBO then proceeds iteratively: First, a regression model, $\mathcal{Q}$, is fitted to the current dataset, $\mathcal{H}$, yielding a surrogate model for $F$ at the current iteration, which may be denoted $\mathcal{Q}_i$. Based on $Q_i$, the next input, $\vec{d}_{i+1}$ to $F$ is selected by optimizing an \textit{acquisition function}, $\alpha:\mathcal{D}\mapsto \mathbb{R}$ over $\mathcal{D}$, which measures the utility gained from evaluating $F$ at $\vec{d}_{i+1}$. Following evaluation of $F\left(\vec{d}_{i+1},\vec{x}_i;\vec{z}\right)$, $\mathcal{H}$, is updated as $\mathcal{H}=\mathcal{H}\cup\left(\vec{d}_i,y_i\right)$. The SMBO process continues until a predetermined time or computation budget is exhausted.

SMBO techniques are typically distinguished by the forms of the surrogate model, $Q$, and the acquisition function, $\alpha$. In Bayesian optimization (BayesOpt), a Gaussian Process (GP, \cite{rasmussen2003gaussian}) is typically used to model a prior over $Q$, which, at each iteration, is updated using previously observed data $\mathcal{H}$ to give a posterior predictive distribution $p\left(y\mid \vec{d},\mathcal{H}\right)$ \cite{shahriari2015taking,Dewancker2015}. Several methods using GPs as surrogate models may be distinguished according to the form of the covariance kernel parameterizing the GP \cite{shahriari2015taking,frazier2018tutorial}. In lieu of GPs, BayesOpt algorithms have also used random forests \cite{Hutter2011} and \textit{tree-based Parzen estimators} (TPE) \cite{Bergstra2011,Bergstra2013} as priors over $Q$.  Acquisition functions are chosen to balance \textit{exploration} and \textit{exploitation} in the sample domain. The most common acquisition function used by these methods is based on an expected improvement criterion \cite{jones1998efficient}; however, newer methods use variations on knowledge gradients \cite{frazier2009knowledge,wu2017bayesian}. 

Many SMBO algorithms can run trials in parallel, which may yield reduced wall clock time (although the total number of trials required may be identical to that used by sequential implementations) \cite{Snoek2012}. One method to further reduce the running time of SMBO \textit{trials} when model evaluations require an inner iterative loop to achieve stationarity (as in BISTRO) is to incorporate an early stopping rule for simulation evaluations that are likely to eventually be extremely suboptimal. An example of such an approach is ``freeze-thaw Bayesian Optimization'' \cite{swersky2014freeze}. 

Recent efforts in transportation science and operations research have also sought to develop tractable simplifications of scenario-based optimization of simulated urban transportation systems. One vein of research concentrates on deriving deterministic analytic equations describing system dynamics at equilibrium from static information, $\vec{z}$ (\textit{e.g.}, network topology and bus schedules) to inform purely functional metamodels~\cite{Osorio2015,Osorio2015a,chong2017simulation}. For example, \cite{osorio2013simulation} combine a computationally tractable model of congested traffic based on queuing theory with a detailed local approximation using a linear combination of basis functions from a parametric family. Alternatively--and analogously to the ``freeze-thaw'' Bayesian optimization setting highlighted above--some approaches use information about the \textit{process} by which the stochastic simulation achieves stationarity to develop techniques that rapidly evaluate different settings of the decision variable vector, $\vec{d}$, while avoiding the need to reach convergence \cite{Cascetta1989,Flotterod2017}.

The present work refrains from prescribing a single best approach to solve optimization program \Cref{eq:constraints,eq:opt_problem_objective}. Instead, the intent of BISTRO is to enable \textit{replicable} future research in this area by providing a platform and problem setting that is generalizable across different planning contexts as well as approachable and of research interest to the ML/AI community. Problem characteristics such as the propensity for competing metrics to be present in system objectives, preemptive stopping of inner optimization loops, the high dimensionality of the search space, and the potential for hybridization of functional and physical metamodels are expected to provide challenging, scalable, and, critically, \textit{explainable} solution approaches. Algorithms combining data-driven dimensionality reduction techniques as well as efficient experiment design can result in repeatable protocols to effectively constrain more general local and global search techniques. Following a presentation of the model architecture, in \cref{sec:pilot_study}, we empirically explore the effectiveness of some of these solution strategies as well as the interpretability of their outputs.

\section{Berkeley Integrated System for TRansportation Optimization (BISTRO)}
\label{sec:architecture}
BISTRO is a new analysis and evaluation platform that works in concert with an ABMS (BEAM) to enable the open-sourced development and evaluation of transportation optimization methods in response to given policy priorities.  This section gives an in-depth description of the BISTRO framework and all of its major components, providing an overview of their purpose, use, and functionality as well as calling attention to the most novel aspects of its design.

\subsection{System Architecture}

\begin{wrapfigure}{R}{0.65\textwidth}
\includegraphics[width=0.65\textwidth]{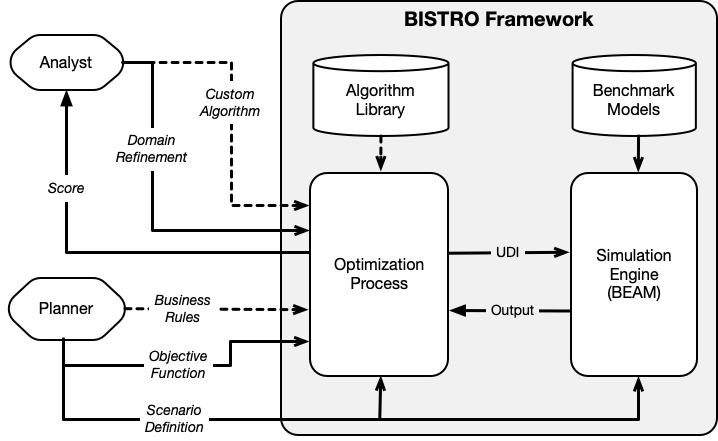}
\caption{BISTRO software architecture, illustrating how the optimization process modulates the flow of information between the BEAM simulation as well the two primary user types. The distinction between the \textit{planner} and the \textit{analyst} is critical in that we do not expect the analyst (an expert in applied ML/AI-based optimization methods) to have transportation or planning background, yet still they should be able to develop generalizable algorithms that can be used to optimize transportation system objectives set by the planning organization.}
\label{fig:bistro}
\end{wrapfigure}

As indicated in \Cref{subsec:simulation-problem}, BISTRO implements elements of the travel demand planning process coupled with components of an automated simulation-based optimization system. This section describes the high-level overall architecture of BISTRO (depicted in  \cref{fig:bistro}), focusing on the conceptual distinction between features relevant to scenario developers and those more appropriate to algorithm designers.

\begin{wrapfigure}{R}{0.5\textwidth}
 \includegraphics[width=0.5\textwidth]{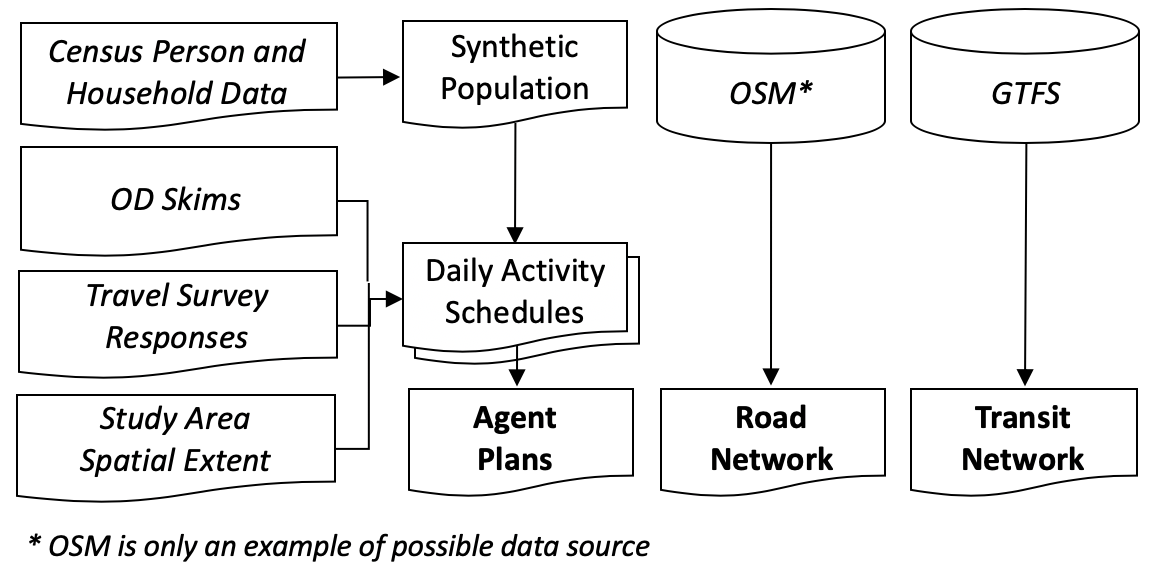}
\caption{Generation of fixed inputs. Italicized entities represent the necessary data to generate fixed inputs (in bold).}
\label{fig:data_activity_synthesis}
\end{wrapfigure}

A BISTRO run environment is configured using a set of fixed input data defining the required transportation system supply elements (\textit{e.g.}, road network, transit schedule, on-demand ride fleet) and demand elements (\textit{e.g.}, synthetic population, activity plans, and mode choice function parameters). Precisely which aspects of the virtual transportation system should be represented in the simulation model depends on the strategic goals and system objectives defined as part of the planning and analysis process motivating a particular BISTRO use case. An example of the set of raw inputs and pre-processing steps is illustrated in \cref{fig:data_activity_synthesis}.

A boundary separates external, exogenously defined inputs from the BISTRO simulation optimization pipeline. Outside of the boundary, the \textit{user-defined inputs} (UDIs) represent the investment, incentive, and policy levers applicable to and available for the study at hand. Concretely, algorithm developers encode solutions as numeric values that represent vector-valued variables controlling aspects of the initialization and evolution of the simulation. For example, a UDI that alters frequency of buses on a route must specify a target transit agency, a route, a start time, an end time, and the desired headway.

While BISTRO maintains a library of available interventions compatible with BEAM, scenario designers, policy makers, and other stakeholders will often want assurance that infeasible, regressive, or otherwise undesirable input combinations are prevented from being selected as ``optimal.'' Together with syntactic and schematic validation of inputs, flexibly-defined \textit{business rules} can effectively act as constraints on the search space---enhancing the interpretability and, thereby, the rhetorical and communicative value of BISTRO-derived solutions.

Just as UDIs from previously conducted BISTRO-based studies are actively maintained and made available to scenario designers, the BISTRO community contributes to a growing library of recommended optimization algorithms that, when evaluated across multiple BISTRO benchmark scenarios, demonstrate desirable performance characteristics. Thus, users lacking resources or expertise to develop optimization routines in-house can still benefit from what, we anticipate, will be cutting-edge research on algorithms and strategies to optimize the simulation of demand-responsive cyberphysical infrastructure.

Project owners of BISTRO deployments may work with stakeholders to develop representative models that will be used to benchmark optimization algorithms. Enabling a well-defined benchmark mechanism permits data on the performance of user-supplied algorithms to be compared. These comparisons can be used to assist in identification of design patterns and computational strategies that advance the state of the art in simulation-based optimization of urban transportation systems.

% While BISTRO is generally amenable to black-box or non-linear optimization approaches, we have found that more effective algorithms create reduced order surrogate models of BEAM using gray- or white-box approaches. This follows from:
% \begin{enumerate}
%     \item the availability of event data collected by BEAM then serialized into useful aggregate outputs via BISTRO-provided processing utilities
%     \item the rich body of theoretical work characterizing human mobility behavior
% \end{enumerate}
% \todo[inline, size=\small, color=blue!40]{I think it's important that this point be emphasized (i.e., "human in the loop algorithmic decision-making"... incidentally a good keyword for this/TRC paper), but this paragraph may be more appropriate in the discussion/conclusion. -SF}
% Compared to black-box algorithms, both gray- and white-box approaches lend themselves more to outcomes that are interpretable from a policy perspective. This is of critical importance to the design of BISTRO, as the results of the optimization framework's evaluations are intended to \textit{inform} rather than \textit{direct} public conversations about proposed policy, investments, and regulations. Consequently, BISTRO encourages constructive collaboration between engineers in the planning community as well as in the ML and AI communities. 

% Put another way: as intelligence functions automating civil infrastructure become increasingly pervasive, BISTRO aims to facilitate more robust and informed community involvement.

\subsection{Scoring Function Design}

Transportation system intervention alternatives are scored in BISTRO based on a function of score components evaluated using \textit{key performance indicators} (KPIs) of the simulation. KPIs for a given simulation should be selected in accordance with the operational, environmental, and social goals, or,  \textit{system objectives} developed as part of the participatory planning process described in \cref{subsec:trans_planning_background}. BISTRO project planners may select  KPIs to include in the scoring function from an existing library of options, or may choose to develop additional KPIs, as appropriate, for the goals and system objectives of the project. Additionally, the form of the scoring function may be designed by the analyst in consultation with the project planner.

% This subsection describes how score components are defined and provides some examples demonstrating compatible functional forms.
\subsubsection{Key Performance Indicators}
\label{sec:KPIs}

\paragraph{KPI overview} There are two general types of KPIs developed in BISTRO: 1) KPIs that measure the operational efficiency of the transportation system (\textit{e.g.}, \textit{vehicle miles traveled} [VMT], vehicle delay, operational costs, revenues) and 2) KPIs that evaluate the experience of transportation system users (\textit{e.g.}, generalized travel expenditure, bus crowding experienced, accessibility). KPIs can be aggregated or disaggregated into score components to support particular policy objectives. For example, the accessibility KPI (detailed below) may be disaggregated by activity type, time period, mode used, and/or sociodemographics in order to evaluate the distributional equity of access provided across different opportunities at varying times of day and/or across population segments of concern.

In practice, any KPI that may be evaluated from the set of output variables (see \cref{sec:Output}) produced by a BISTRO simulation run may be included as a score component in the scoring function. However, careful consideration of candidate KPIs must include an evaluation of the sensitivity of the metric to the UDIs of interest as well as the efficiency of the KPI in providing the desired feedback regarding the optimality of outcomes of alternative UDI values. For example, \textit{person miles traveled} (PMT) is a commonly used metric in transportation system performance measurement to gauge the amount of mobility delivered by the system. Yet, PMT is highly invariant within a scenario in BISTRO due to the fact that agent plans are fixed. Thus agents will make the same trips regardless of the UDI values and the miles traveled by each agent will only vary in so much as the networks available for each mode offer more or less direct paths to travel from the origin to destination of each trip.\footnote{For example, a transit mode choice for a particular trip may result in more PMT than a walk mode choice for the same trip, as the sidewalk network may enable a more direct path.}

% \textit{Modal split}, the distribution of transportation modes used across a set of trips, is another commonly used metric that often serves as an indicator of the sustainability of the distribution of demand across available modes in a transportation system. Reduction of the modal split of single occupant vehicle use, for example, is often used as a goal for meeting an objective related to reducing congestion and/or GHG emissions relative to the quantity of trips made. However, BISTRO enables the measurement of more precise goals that directly measure the system objectives. Rather than including the modal split of single occupant vehicle trips in the scoring function, one could use a direct measure of vehicle occupancy, delay experienced, and/or GHG emissions produced in the scoring function. Nevertheless, the BISTRO visualization suite (see \cref{sec:Output}) enables the user to analyze the relationship of the optimized UDI outcomes based on a particular scoring function with metrics such as modal split that provide additional intuition of the aggregate system impacts.

\paragraph{Implemented KPIs} The following items represent categories of KPIs that have been developed and implemented in BISTRO at the time of publication of this article:
\begin{enumerate}
    \item \textit{Accessibility.} 
    %While the term accessibility takes on a variety of meanings in different contexts, 
    In an urban transportation planning setting, accessibility has often been defined as a measure of the ease and feasibility with which opportunities or points of interest can be reached via available modes of travel. 
    %Although there are many ways to measure accessibility, 
    It is quantified in BISTRO as the sum of the average number of points of interest (of a specific type of activity) reachable within a given duration of time, with functionality also provided to measure mode-specific accessibility.
    %as the sum of the average number of points of interest reachable from network nodes by car or using public transit, within a specified amount of time during specific time periods.
    
    \item \textit{Trip Expenditure and Generalized Transportation Cost Burden.} The socio-demographic and spatial heterogeneity of travel behavior within BISTRO enables a variety of equity-focused impact analyses. Two such metrics have been implemented in BISTRO: average trip expenditure and average generalized transportation cost burden. While the former is the average monetary cost incurred by agents per trip, the latter is computed as the sum of the travel expenditures of the trip (costs of fuel and fares minus incentives, as applicable) and the monetary value of the duration of the trip (the product of the total trip duration and the population \textit{average value of time} (VOT)), divided by the household income of the agent completing the trip. the monetary value of the trip duration is calculated by multiplying total duration by the population average VOT. Both KPIs may be disaggregated to emphasize particular equity goals (e.g., by socio-demographic groups, trip purpose, mode, etc.).

    \item \textit{Bus Crowding.} The \textit{level of service} (LoS) experienced by public transit passengers has a direct influence on short- and long-term demand for public transit service. 
    %In the short-term, passenger demand for a particular transit line is dependent on the time and cost of alternative travel options. Thus, the frequency and service period of transit service determines the availability, wait times, transfer times, and in-vehicle times of a prospective transit trip and thus the utility of that trip in comparison to the same trip completed with alternative transportation modes. In addition, the available capacity on a transit vehicle affects whether or not the passenger can board transit at the desired time. Furthermore, the comfort afforded by the available space on a transit vehicle has long-term effects on transit demand, as passengers internalize their experience during many transit trips over time and develop an additional aversion or affinity to transit based on their expectation of the LoS. Upon experiencing the discomfort of an overcrowded transit vehicle for the same ‘trip’ (\textit{e.g.}, a traveler’s 8 am home--work morning commute trip), a traveler will come to expect that LoS when considering whether to take transit for that trip in the future. 
    In addition to cost and travel time factors, the available capacity on a transit vehicle affects whether or not a passenger can board a public transit vehicle at their desired time as well as the level of comfort they experience during the trip. Though the LoS of public transit may be measured in BISTRO by any one of the factors mentioned, BISTRO includes a ready-made example of an LoS KPI related to passenger comfort: average bus crowding experienced. This metric is computed as the average over all transit legs of the total passenger-hours weighted by VOT multipliers corresponding to the load factor (the ratio of total passengers to the seating capacity) of the bus during the leg. 

    \item \textit{Vehicle Miles Traveled (VMT) and Delay}. The BISTRO KPI library includes three examples of congestion score components that provide insight into the destination- or opportunity-independent level of mobility on a network, the overall network performance, and efficiency: total VMT by all motorized vehicles in the transportation system, total vehicle delay, and average vehicle delay experienced per passenger trip. Total vehicle delay is calculated as the sum over all path traversals of the difference between the realized duration and the free flow travel time of the traversal. Vehicle delay experienced per passenger trip is calculated as the total difference between the realized duration and free flow travel time of all legs of a trip completed by modes subject to congestion.  

    \item \textit{Financial sustainability.} Most system interventions will have some impact on the flow of funds in or out of the transportation system. Including a KPI that helps stakeholders understand the general financial impacts of such interventions is necessary. The financial sustainability metric provided in the BISTRO KPI library is the sum of all public transit fares collected minus all incentives distributed (if any) and all operational costs of the public transit system \footnote{The operational costs include the total costs of fuel consumed, and hourly variable costs of bus operations (see \cref{table:vehTypes} for an example of operational costs). Hourly variable costs include estimated labor, maintenance and operational costs. The rates for each of these factors is specified in the vehicle fleet configuration variables.}. In the event that a BISTRO project does not alter public transit service, the operational costs may be omitted from the KPI, if desired. 
    %The purpose of the financial sustainability KPI is to shed insight on the tradeoffs from varying transit LoS while varying incentives for transit and other modes. 
    
    \item \textit{Environmental sustainability.} The environmental sustainability of a transportation system intervention may be measured as the local and/or global impacts to the system. In addition to VMT- and fuel efficiency-based estimates, BEAM enables estimation of emissions directly from the simulated fuel consumption, based on the realized speeds traveled by each vehicle throughout a simulation run\footnote{For more information on the methodology followed to estimate fuel consumption, please refer to the BEAM documentation \url{https://beam.readthedocs.io/en/latest/index.html}.}. The VMT-based fine particulate emissions (PM$_{2.5}$) KPI captures local environmental sustainability via a mileage-based measure of  air quality impacts based upon vehicle type.
    %total PM$_{2.5}$ emissions produced by all motorized vehicles during the simulation. Using criteria pollutants, specifically particulate matter running exhaust emissions factors, provides . 
    Additional local emissions KPIs may easily be included using the appropriate emissions factors\footnote{For more information on the methodology followed to develop this metric, please refer to the California Air Resources Board documentation, \url{https://www.arb.ca.gov/cc/capandtrade/auctionproceeds/cci_emissionfactordatabase_documentation.pdf}.}. A \textit{greenhouse gas} (GHG) emissions KPI allows the optimization to explicitly account for fuel-consumption-based global environmental sustainability \footnote{It is important to note that the GHG emissions KPI will be correlated with VMT and fine particulate emissions. Thus, inclusion of all three KPIs creates a suite of environmental sustainability metrics that may apply disproportionate weight on environmentally-related objectives, which may or may not be desirable for certain policy agendas. Project planners may choose to apply scaling factors (as described in \cref{subsubsec:ScoringFunction}) to balance the influence of the environmental sustainability score components.}.
\end{enumerate}

\subsubsection{Scoring Function}
\label{subsubsec:ScoringFunction}

\begin{wrapfigure}{R}{0.55\textwidth}
 \includegraphics[width=0.55\textwidth]{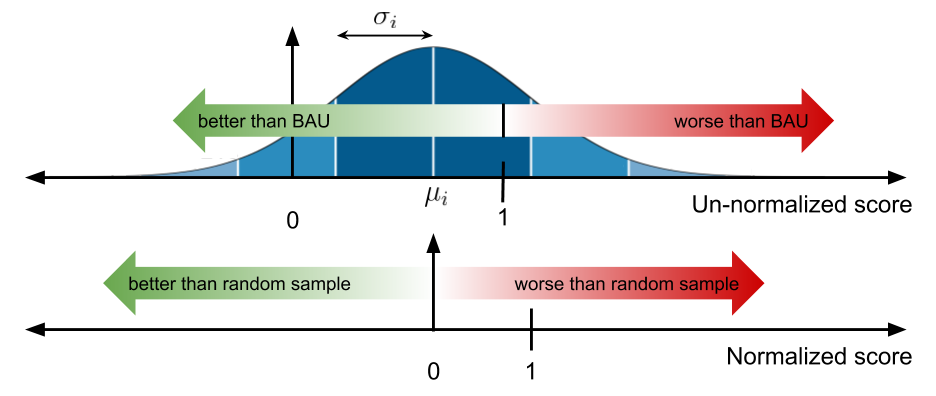}
    \caption{A visual representation of the normalization procedure for a hypothetical score component, $i$. The ratio of the submission score and BAU score (depicted in the upper plot) is normalized by taking its $z$-score (depicted in the lower plot) relative to a random input sample.}
    \label{fig:score_normalizing}
\end{wrapfigure}

The BISTRO scoring function serves as the objective function by which the UDIs are optimized. The selection and/or definition of the objective function in accordance with the project directives is considered to be the responsibility of the project planner. Herein, a general structure is defined to facilitate the creation of custom objective functions. Multiple project objectives (referred to here as \textit{score components}) may be included in the scoring function--either as individual elements within a vector of scalar-valued score components to be minimized, or as parameters to a function that aggregates the objectives into a one-dimensional scalar score. The score components are computed as the normalized ratio of the value of the corresponding KPI in the given simulation run to the value of the same KPI in the \textit{business-as-usual} (BAU) run\footnote{In the BAU of a given scenario, the simulation is run without alteration from the initial configuration of that scenario.}. The improvement ratios are normalized using KPI values produced by a randomized sample of the UDI space, the size of which can be defined by the BISTRO project owner. This normalization (depicted graphically in \cref{fig:score_normalizing}) accounts for differences in variance across KPIs, thus allowing the score components to provide meaningful feedback on the improvement achieved for each KPI relative to the distribution of the ratios of KPI to BAU produced by the random search. The composite score is thus a function of the normalized relative improvements of the candidate input to the BAU in each metric, as follows:

    \begin{equation}
        F\left(\vec{C}_{s}, \vec{K}, \vec{\sigma},\vec{\mu}, \vec{\alpha}\right) = f\left(\vec{z},\vec{\alpha}\right),
    \end{equation}
where $\vec{K}$ is the vector of all KPIs evaluated for a given set of inputs, $\vec{C}_s$; $\vec{\mu}$ and $\vec{\sigma}$ are the vectors of normalization parameters; and $\vec{z}$ is a vector of each KPI's $z$-scores, \textit{i.e.},
\begin{equation}
\label{eq:zscore}
    z_i = \frac{\frac{K_i(C_s)}{K_i(C_{BAU})} - \mu_i}{\sigma_i},
\end{equation}
for the $i$-th KPI. The value of the $i$-th score component in the BAU case is simply $K_i(C_{BAU})$.

The default objective is to minimize the composite score function, since an increase in many of the score components actually represents a scenario that is worse than the \textit{status quo} (\textit{e.g.}, decreasing VMT over BAU results in a lower unscaled score than increasing VMT). To maintain consistency in this regard, the scoring function may include an additional parameter $\vec{\alpha}$ to allow for transformation of score components that are positively related to desirable outcomes (\textit{e.g.}, improvements in accessibility). For example, if the scoring function takes the form of a sum over all score components, the parameter $\vec{\alpha}$ may be used as a coefficient of each score component that determines whether the component will be summed or subtracted, as follows:
\begin{equation}
    \alpha_i = \begin{cases}-1 \ \text{if it is desirable for score component $i$ to increase} \\
    1 \ \text{otherwise}
    \end{cases}
\end{equation}

% \begin{figure}[h]
%     \centering
%     \includegraphics[width=0.6\textwidth]{figures/score_normalizing.png}
%     \caption{A visual representation of the normalization procedure for a hypothetical score component, $i$. The ratio of the submission score and BAU score (depicted in the upper plot) is normalized by taking its $z$-score (depicted in the lower plot) relative to a random input sample.}
%     \label{fig:score_normalizing}
% \end{figure}

This approach is distinct from the one typically used for \textit{Cost-Benefit Analysis} (CBA) tasks in urban planning practice, in that it seeks to optimize an aggregate function of the relative improvements in each KPI rather than optimizing the net improvement from all KPIs. While CBA often draws skepticism due to the discretion inherent in the process of converting all KPIs into a common unit such as time or money so that the net value of costs and benefits can be computed, the approach taken in the BISTRO scoring function does not require any such assumptions to be made. Rather, each score component represents the relative improvement over the BAU that is achieved by a simulation run using a particular set of UDIs. Objective function designers may choose to apply additional scaling factors to the score components using the $\vec{\alpha}$ parameter vector.

% \begin{itemize}
%     \item considerations for choosing a set of components: sensitivity to inputs, correlation between components (can choose correlated components on purpose to emphasize particular input-output relationships), ability to directly measure goals (e.g., don't use mode share, use the metric(s) for which mode share is typically used as a proxy - exception: active mode share),
    
% \end{itemize}

\subsection{Inputs}
% \subsubsection{Preparation of Fixed Inputs}
\paragraph{Preparation of fixed inputs} For each BISTRO study, a set of fixed inputs must be provided to BEAM. For a given study area, these typically include the road network, the transit schedule, and the demand profile. Depending on the system objectives, additional data may be necessary to fully configure the simulation. \cref{fig:data_activity_synthesis} illustrates a schematic of the inputs for a typical simulation.
% \begin{figure}[ht]
% \centering
% \includegraphics[scale=0.4]{figures/data_population_synthesis.png}
% \caption{Generation of fixed inputs. Italicized entities represent the necessary data to generate fixed inputs (in bold).}
% \label{fig:data_activity_synthesis}
% \end{figure}

The \textit{road network}, including the physical properties of its links and nodes, may be generated using \textit{Open Street Maps} (OSM) data for the geography of interest. The \textit{transit network} configuration follows the  easily accessible \textit{General Transit Feed Specification} (GTFS) format.  \textit{On-demand ride services}   (\textit{Transportation  Network  Companies}  [TNCs]  such as Uber, Lyft, and Via), are modeled as a fleet of vehicles driven by agents that are exogenous to the population, or may be driven autonomously. The initial locations of the vehicles may be sampled randomly or from a specified distribution in accordance with appropriate data.
% On-demand ride vehicles undergo three phases of service: empty, fetch (the vehicle is reserved and is en-route to pickup a passenger) and fare (a passenger is in the vehicle traveling to its requested destination).
The price of on-demand rides is fixed, consisting of a distance-based and a duration-based component. The size of the on-demand ride service fleet is a proportion of the total number of agents in the simulation, as determined by a configuration parameter. Driver repositioning behavior when not currently driving to or serving a passenger can be configured to follow one of several repositioning algorithms defined within BEAM.
% Each on-demand driver agent begins each iteration of a simulation run located a random distance from the \textit{household} of a selected agent. Several repositioning algorithms are available in BEAM to specify how agents behave when not carrying passengers. The default behavior causes driver agents who have just dropped off a passenger to stop driving and wait at the drop-off point until receiving a subsequent request.

% The transit network is a subgraph of the road network comprised of a set of transit facilities (transit stops) and a set of routes. 

At the start of the simulation, a \textit{synthetic population} of virtual agents and households is generated such that the sociodemographic attributes of these virtual entities are spatially distributed in accordance with real-world census and/or location-based data from the city of interest. Each agent follows a daily plan consisting of several activities throughout the day. As illustrated in \Cref{fig:data_activity_synthesis}, these \textit{daily activity schedules} are generated based on \textit{origin-destination} (OD) skims (matrices that provides the number of trips between zones), travel surveys, and zonal boundary spatial data. 

% \subsubsection{Calibration} 
\paragraph{Calibration} Prior to use in optimization runs, BEAM needs to be calibrated to empirical data by iteratively adjusting model parameters until a simulation outputs representing traffic patterns match their real-world counterparts with minimal error\footnote{BEAM calibration is typically targeted at mode split, volumetric traffic counts, and travel distance distributions. The choice of which target(s) to use may depend on regulatory requirements, literature recommendations, or precedent \cite{Zheng2013,bucci2018fhwa}.}. Calibration of BEAM for usage in BISTRO should be limited to adjustments of behavioral parameters controlling microscopic decision-making (e.g., mode choice intercepts, prices, regulation-driven incentives/tariffs)\footnote{Often, due to computational constraints, a sub-sample of a full population is simulated. The capacity of physical resources (e.g., road network link carrying capacity, maximum transit occupancy, number of electric vehicle charging plugs per station) may need to be adjusted based on the size of the population sample. For evaluation purposes, the outputs of a sub-sampled simulation are often scaled back up to the full population.}. While it is possible to adjust many additional BEAM parameters to reduce calibration error, this practice should be discouraged, as it may result in models that are overfitted to a state of the world represented by a particular ground-truth dataset, thereby limiting the calibrated simulation model's use in predictive contexts. In addition to the representativeness of ground truth data, the quality and quantity of input data (\textit{e.g.}, network data resolution or population spatial resolution), may influence the extent to which the model is able to achieve calibration end points. 

% \subsubsection{Configuration of User-Defined Inputs}

\begin{wraptable}{R}{0.5\textwidth}
\caption{Example of bus frequency adjustment input file.}
\label{table:example_input_frequency}
    {\footnotesize
    \begin{tabular}{ccccc}
        \toprule
         route\_id & start\_time & end\_time & headway\_secs & exact\_times \\
         \midrule
         1340 & 21600 & 79200 & 900 & 1\\
         1341 & 21600 & 36000 & 300 & 1\\
         1341 & 61200 & 72000 & 300 & 1\\
         \bottomrule
    \end{tabular}
    }
\end{wraptable}

% \begin{table}[H]
%     \caption{Example of bus frequency adjustment input file.}
%     \centering
%     {\footnotesize
%     \begin{tabular}{ccccc}
%         \toprule
%          route\_id & start\_time & end\_time & headway\_secs & exact\_times \\
%          \midrule
%          1340 & 21600 & 79200 & 900 & 1\\
%          1341 & 21600 & 36000 & 300 & 1\\
%          1341 & 61200 & 72000 & 300 & 1\\
%          \bottomrule
%     \end{tabular}
%     }
%     \label{table:example_input_frequency}
% \end{table}
\paragraph{Configuration of UDIs} BISTRO provides a library of possible inputs for scenario designers to adapt to specific use cases. The selection of UDIs is intended to be compatible with the system objective. UDIs may represent, for example, the investment (\textit{e.g.},  transit fleet mix modification, bus route modifications, parking supply, electric vehicle charge station locations, dynamic redistribution of e-bikes or on-demand vehicles), incentive (\textit{e.g.}, incentives to specific socio-demographic groups for selected transportation modes, road pricing/toll roads, fuel tax), or policy/operational (\textit{e.g.}, transit schedule adjustment, transit fare modification, parking pricing) levers applicable to the study at hand. The project owner may constrain the range of possible values upon which each UDI is valid by setting the corresponding input validation parameters and business rules. The example input file for bus scheduling shown in \Cref{table:example_input_frequency} defines alteration of the headway of a particular bus route during a particular service period (defined by its start and end times).

\subsection{Output Analysis and Visualization}
\label{sec:Output}

% The outputs of BEAM simulations allow stakeholders to better understand the implications of proposals identified using optimization algorithms. For example, visualizations of congested roadways with millions of agents behaving independently can provide a concise method to communicate the effects of infrastructure interventions. 

 The \textit{raw} outputs of a BEAM simulation include millions of events that reflect the microscopic actions of each agent as they make their way through the day. While this is a detailed history of what transpired during the day for each agent, it does not provide planners with explanatory insight into how perturbations of input variables influence model behavior leading to changes in output responses.  To simplify exploration of alternatives to purely black box optimization methods, the BISTRO platform provides a suite of tools that process and organize event data from BEAM simulation runs into a relational format, permitting relationships between each person’s activities, trips, path traversals to be queried at different levels of detail\footnote{Further details on the output of the parser including an entity relationship diagram are available at \url{http://bistro.its.berkeley.edu/assets/download/pdfs/General_System_Specification.pdf}}. A Jupyter notebook then incorporates the post-processed outputs into a standard suite of multivariate analyses and visualizations, thereby facilitating interpretation and communication of the effect that policies have on system objectives.

\subsection{Implementation Details and Performance Characteristics}

Both BISTRO and BEAM are primarily implemented in Scala. Input files are read from a single directory and injected into the BEAM initialization routine. The system is containerized using Docker, which helps to facilitate OS-agnostic local and remote execution.

The runtime of BEAM depends on various inputs including population size, network resolution, transit network, ridehail fleet size as well as available compute resources such as number of processors, memory, etc. On a machine with 32 (2.5GHz Intel\textregistered  Xeon\textregistered Platinum 8175) CPUs and 128GB RAM, the runtime for a 315,000 agent simulation of a San Francisco Bay Area scenario (representing a 25\% sample of the approximate 2018 population) takes around 14 hours to complete 15 iterations. On a machine with identical compute resources, a 15,000 agent simulation for the the Sioux Faux scenario (see \cref{sec:sioux_faux}) takes approximately 30 minutes to complete 15 iterations.

Currently, the primary performance bottleneck in BEAM is routing. The routing engine generates millions of routes (reflecting multimodal options for agents to choose between) for a single simulation run. Some additional overhead considerations such as data availability, level of model resolution required, as well as the impact of augmented BEAM functionality must be balanced in light of available computational resources.

\section{Initial Pilot Study and Launch}
\label{sec:pilot_study}
\begin{figure}[h!]
\begin{center}
 \includegraphics[width=\textwidth]{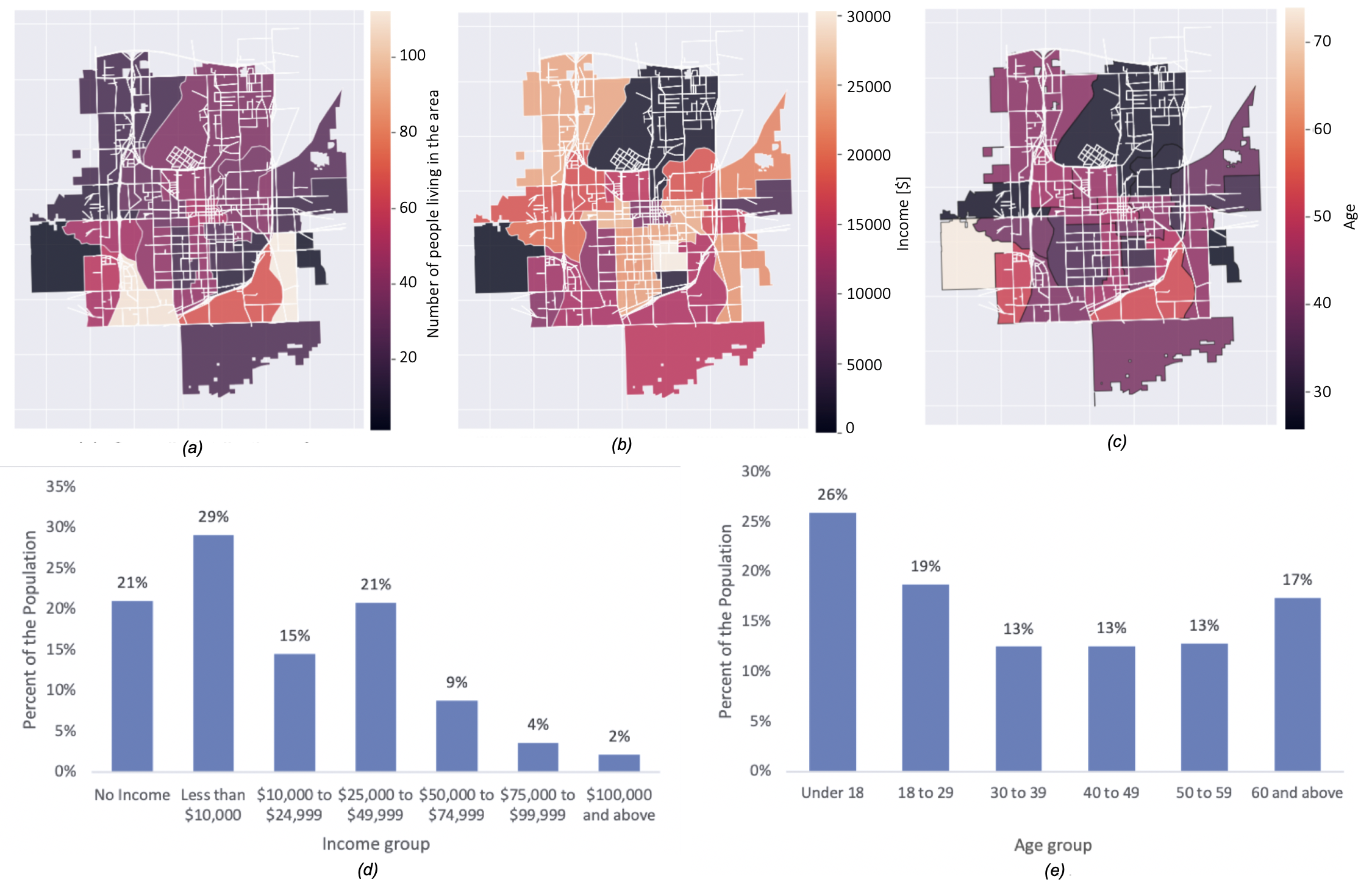}
\caption{Demographics of Sioux Faux. (a) Overall distribution of the population per census tract. (b) Distribution of the median population income per census tract. (c) Distribution of the median population age per census tract. (d) Overall population income distribution. (e) Overall population age distribution. \label{f:demographics}}
 \vspace{-0.2in}
\end{center}
\end{figure}

\subsection{Sioux Faux}
\label{sec:sioux_faux}

% \textit{This subsection describes the toy city scenario. It will touch on the population and plans synthesis methodology, as well as the implementation of transportation modes, network, etc.}
An agent-based model of transport supply and demand inspired by the real city of Sioux Falls, South Dakota\footnote{The ``Sioux Falls'' scenario is a commonly used benchmark in ABMS research, see \url{https://github.com/bstabler/TransportationNetworks/tree/master/SiouxFalls}} was adapted for the purpose of developing and testing example scenarios within BISTRO. To underscore that for these purposes, such scenarios were not developed to be true replicas of the city of Sioux Falls, this benchmark BISTRO scenario is referred to as \textit{Sioux Faux}. The scenario configuration, input specification, and scoring function were designed to support strategic objectives of financial and environmental sustainability, reduced congestion, and improved equity, accessibility, and transportation system level of service. 

\begin{wrapfigure}{R}{0.5\textwidth}
  \includegraphics[width=0.5\textwidth]{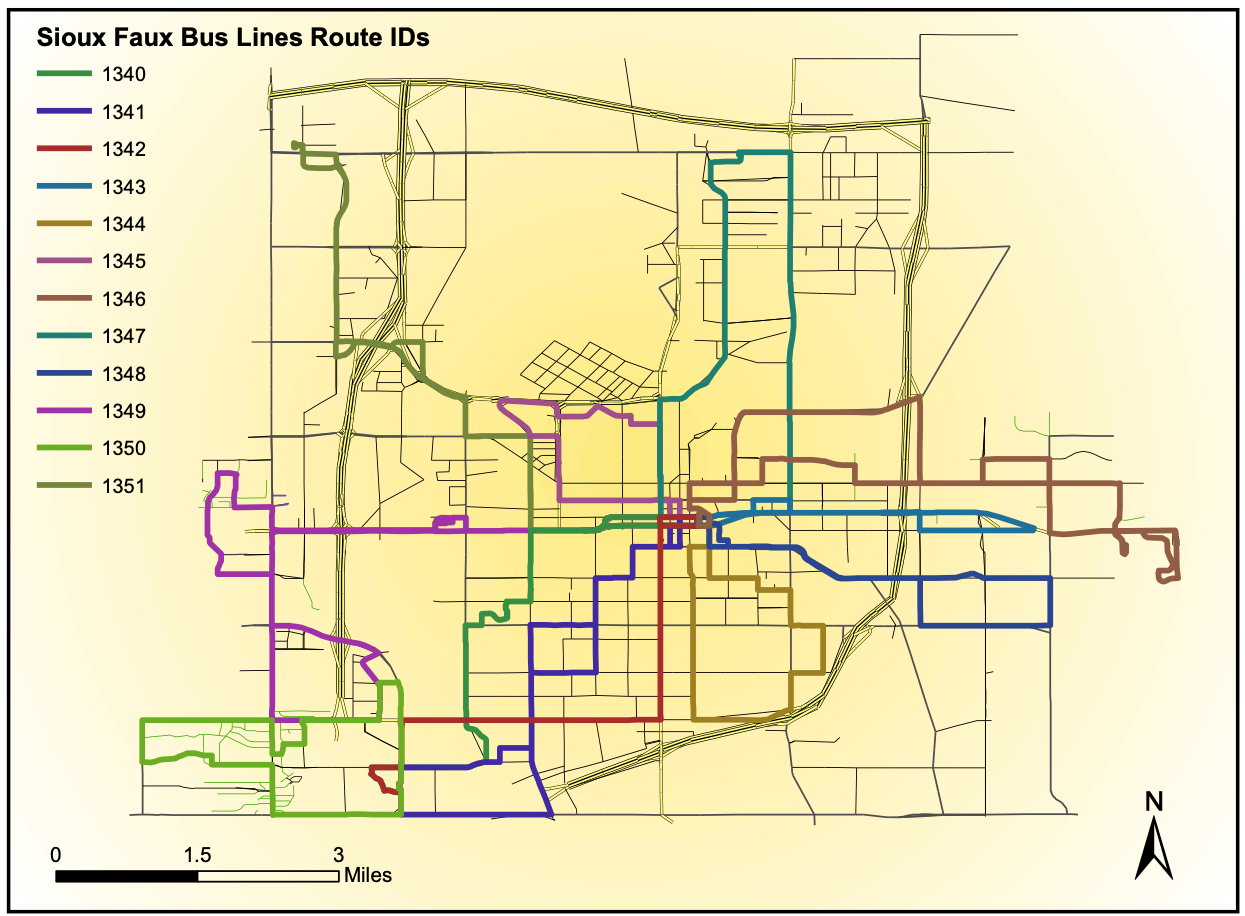}
    \caption{Sioux Faux bus and road networks.}
    \label{fig:bus_map}
    \vspace*{-0.15in}
\end{wrapfigure}

\subsubsection{Scenario Configuration}
\paragraph{Population and plan synthesis} The synthetic population of Sioux Faux was generated using publicly-available census data for the city of Sioux Falls, South Dakota as inputs to the Doppelganger library\footnote{Doppelganger uses a novel Bayesian optimization approach combined with the hierarchical list-balancing algorithm developed as part of the PopSyn library \cite{Vovsha2015}. For more information about the Doppelganger library, see \url{https://github.com/sidewalklabs/doppelganger}}, a state-of-the-art population synthesis framework developed in Python.  Specific inputs to Doppelganger used to generate the Sioux Faux population included household and individual \textit{Public Use Microdata Sample} (PUMS) data for South Dakota from the 2012-2016 (5-year) \textit{American Community Survey} (ACS), which is conducted annually by the US Census\footnote{The 5-year PUMS comprises a 5\% sample of the US population. It is computed as an aggregate of 1-year samples, which themselves aim to survey 1\% of the US population}. The \textit{Public Use Microdata Area} (PUMA) for Sioux Falls constrains the state-wide survey data to our general area of interest. Population demographics derived from the synthetic population for Sioux Falls are shown in \cref{f:demographics}.

An existing set of agent plans for Sioux Falls previously developed for MATSim simulations was used as the basis for the plans of our expanded Sioux Faux population\footnote{More specifically, we modified the Sioux Falls 2016 scenario developed by H{\"o}rl (\cite{Horl2016}), which is an update of a scenario prepared in 2014 by Chakirov and Fourie \cite{chakirov2014enriched}. For more information on the Sioux Falls scenario, see \url{https://www.ethz.ch/content/dam/ethz/special-interest/baug/ivt/ivt-dam/vpl/reports/901-1000/ab978.pdf}}. After initial pilot testing to determine trade-offs between population size, behavioral realism, and computational complexity, we took a 15\% sub-sample of the full synthetic population (approximately 15,000 agents). We used a spatially-constrained sampling mechanism in order to allocate plans to agents in accordance with household locations and census tract household and individual attribute distributions. The subsampling mechanism also enforces realistic constraints on agent plans and behavior using predicates such as ``agents under the age of 18 should not have a work activity'' and ``agents under the age of 16 should not be allowed to drive''. % used to clear float (SAF, 7/7/2019)

\paragraph{Transportation Network} The Sioux Faux transportation network includes a road network accessible to walking agents, personal vehicles, on-demand ride services (TNCs such as Uber and Lyft), and public buses providing fixed-route service\footnote{The initial bus route scheduling is directly generated from the publicly available GTFS for Sioux Falls, which includes erratic headways across routes.}. The on-demand ride services implemented in this scenario include only single-passenger rides (\textit{e.g.}, UberX, Lyft Classic) from a fleet of on-demand ride vehicles that was distributed randomly across the road network at the start of each simulation run. Driving alone is the most frequently used mode, comprising approximately 75\% of the miles traveled for the BAU scenario.

\subsubsection{User-Defined Input Specification}
\label{subsubsec:UDIs_SiouxFaux}
For this initial pilot study, a set of four UDIs were investigated: 
\begin{enumerate}
    \item bus fleet vehicle composition
    \item bus service frequency
    \item bus fare
    \item multimodal incentive program for on-demand rides and public bus trips
\end{enumerate}

In the bus fleet for the BAU scenario, all vehicles were set to a default bus type. Optimization of the bus fleet vehicle composition and service frequency offers the opportunity to improve the level of bus service by better matching the bus type with specific demand characteristics of each route. Four types of buses (including the default) were considered (see \cref{table:vehTypes}), each with different technical properties (seating and standing capacity) and cost characteristics (cost per hour, cost per mile, fuel type and fuel consumption rate).

\begin{wraptable}{R}{0.65\textwidth}
\caption{Values used for $\alpha_i$ in each of the subsequent results sections. For score components that are positively related to desirable outcomes, negative $\alpha_i$ is provided to transform it consistent with a minimization problem.}
    \label{table:kpis}
    {\footnotesize
    \begin{tabular}{llccc}
    \toprule
    KPI & KPI type & \rotatebox{90}{Contest} & \rotatebox{90}{Post-Contest~} & \rotatebox{90}{New KPIs} \\
    \midrule
    accessible work locations & Accessibility & -1 & -1 & -- \\
    accessible secondary locations & Accessibility & -1 & -1 & -- \\
    accessible work locations by car & Accessibility & -- & -- & -1 \\
    accessible secondary locations by car & Accessibility & -- & -- & -1 \\
    accessible work locations by transit & Accessibility & -- & -- & -1 \\
    accessible secondary locations by transit & Accessibility & -- & -- & -1 \\
    average trip expenditure-work & LoS & 1 & 1 & -- \\
    average trip expenditure-secondary & LoS & 1 & 1 & -- \\
    average travel cost burden-work & Equity & -- & -- & 1 \\
    average travel cost burden-secondary & Equity & -- & -- & 1 \\
    average bus crowding experienced & LoS & 1 & 1 & 1 \\
    total vehicle miles traveled & Congestion & 1 & 1 & 1 \\
    average vehicle delay per passenger trip & Congestion & 1 & 1 & 1 \\
    costs and benefits & Financial Sustainability & -1 & -1$^\ast$ & -1$^\ast$ \\
    total grams PM$_{2.5}$ emitted & Environmental Sustainability & 1 & 1 & 1 \\
    total grams GHG$_e$ emitted & Environmental Sustainability & -- & -- & 1 \\
    \bottomrule
    $^\ast$fixed KPI post-contest
    \end{tabular}
    }
\end{wraptable}

A UDI was implemented to vary the bus schedule on each route, including the hours of service and the headway, or service frequency as shown in \Cref{table:example_input_frequency}. Multiple service periods with varying headways on the same route were thus possible. The bus fare UDI allowed for the optimization of the fare on each route, segmented by passenger age groups. Finally, a multimodal incentive UDI was implemented to enable reimbursement for on-demand rides, walk to/from transit, or drive to/from transit trips to qualifying  individuals  based  on  age, income, or both.

\begin{wraptable}{r}{0.45\textwidth}
 \caption{Transit vehicle types available for Sioux Faux bus fleet: (a) Fuel type, (b) Fuel consumption rate ($\mathrm{J}/\mathrm{m}$), (c) Operational cost ($\mathrm{USD}/\mathrm{hr}$), (d) Seating capacity, (e) Standing capacity.}
    {\footnotesize
            \begin{tabular}{clllll}
                \toprule
                \textbf{Vehicle type, }$\bm{ c \in C}$ & a & b & c & d & e\\
                \midrule
                BUS-DEFAULT & diesel & 20048 & 89.88 & 37 & 20 \\
                BUS-SMALL-HD & diesel & 18043.2 & 90.18 & 27 & 10 \\
                BUS-STD-HD & diesel & 20048 & 90.18 &  35 & 20 \\
                BUS-STD-ART & diesel & 26663.84 & 97.26 & 54 & 25 \\
                \bottomrule
            \end{tabular}
        }
    \label{table:vehTypes}
\end{wraptable}

\subsubsection{Business Rules} In order to ensure that optimal solutions would be compliant with common policy and planning practices, four business rules were implemented: 1) there may be no more than five distinct bus service periods (this mimics a typical delineation of transit service provision: am peak, midday, pm peak, evening, late night/early morning), 2) bus route headways may be no more than 120 minutes and no fewer than 3 minutes, 3) bus fares and mode incentives may not isolate a single age, and 4) ages for both fares and incentives may be specified in segments no smaller than five years in range and income for incentives may be assigned in segments no smaller than \$5,000 in range.

% \begin{figure}[ht!]
%     \centering
%     \includegraphics[scale=0.5]{figures/Bus_map.png}
%     \caption{Sioux Faux bus map.}
%     \label{fig:bus_map}
% \end{figure}

\subsubsection{Scoring Function Design}
The set of Sioux Faux UDIs have varying interconnected impacts on the operation of and access to public transit and on-demand ride service by agents. Thus, the scoring function upon which the inputs were optimized was designed to include a variety of metrics that relay feedback on the user experience and operational efficiency of the transportation system as a whole. \cref{table:kpis} presents all of the KPIs used in Sioux Faux scenarios referenced in this text\footnote{Note that several of the KPIs in this table refer to two post-contest follow-on studies, see \cref{subsec:results}}.

Five KPIs were developed to represent three main aspects of user experience: accessibility, travel expenditure, and transit passenger comfort. The accessibility and travel expenditure were both disaggregated by trip purpose such that score components for accessibility and travel expenditure to work and secondary activities were each included separately in the scoring function. Transit passenger comfort was measured as the average bus crowding experienced by bus passengers\footnote{Average bus crowding in the Sioux Faux scenario was calculated as the average number of agent hours spent per transit trip in buses occupied above their seating capacity. This KPI has since been updated, see \cref{sec:KPIs}}.

Four KPIs of operational efficiency were included to account for the congestion, environmental sustainability, and financial sustainability resulting from optimized inputs. Total VMT was included as a KPI for overall congestion while average vehicle delay per passenger trip served as a KPI of the average impact of congestion. The total amount of PM$_{2.5}$ emitted served as a KPI of the environmental impact resulting from each simulation run. Finally, the financial sustainability KPI was included to incentivize outcomes with minimal impact to the bottom line of the transit agency by taking into account the operational costs, incentives distributed, and revenues collected from any combination of transit fleet mix, scheduling, fare structure and incentive program. 

All metrics are aggregated according to the following function:
     \begin{equation}
     \label{eq:zscore_siouxfaux}
        F\left(\vec{C}_{a}, \vec{F}, \vec{\sigma},\vec{\mu}, \vec{\alpha}\right) = \sum_{i \in \mathbb{K}}  \frac{\left(\frac{K_i(C_{s})}{K_i(C_{BAU})}\right)^{\alpha_i} - \mu_i}{\sigma_i}  
    \end{equation}
where all variables are defined as described in \cref{subsubsec:ScoringFunction}, with the set of KPIs and corresponding $\alpha_i$ values as specified in \cref{table:kpis}. We executed 800 runs using randomly generated values of $\hat{d}$ to produce the normalizing statistics (i.e., $\mu_i$s and $\sigma_i$s in \cref{eq:zscore}) for each metric. 

% \[\mathbb{K}:=
% \left\{\!\begin{aligned}
%     &\text{accessibility-work,}\\ 
%     &\text{accessibility-secondary,}\\ 
%     &\text{LoS: average trip expenditure-work,}\\ 
%     &\text{LoS: average trip expenditure-secondary,}\\ 
%     &\text{LoS: average bus crowding experienced,}\\ 
%     &\text{congestion: total VMT,}\\ 
%     &\text{congestion: average vehicle delay per passenger trip,}\\ 
%     &\text{financial sustainability,}\\
%     &\text{environmental sustainability}
% \end{aligned}\right\}
% \]

% Finally, the values of $\vec{\alpha}$ were determined as follows:
% \[\alpha_i = \begin{cases} -1 \ \text{if $i \in$ \{\text{accessibility-work, accessibility-secondary, financial sustainability}\}} \\  1 \ \text{otherwise}
% \end{cases}
% \]

\subsection{Pilot study results}
\label{subsec:results}
% \subsection{Contest Participation (0.5 page)}
% \subsubsection{Contest participation}

\paragraph{Contest participation and results} Over the course of 17 days, 487 people in teams of one to four (mostly consisting of engineers and data scientists with little to no domain expertise in transportation planning) effectively created nearly 1,000 different ``city transportation plans'' for the Sioux Faux scenario consisting of the UDIs described in section \cref{subsubsec:UDIs_SiouxFaux} \footnote{Uber does not endorse any of the solutions presented.}.

To be able to compare their results and scores with other participants, each team could submit up to five solutions per day and thus be ranked in a web-accessible leaderboard. While contestants trained algorithms online, final evaluation, leaderboard, and discussion boards were hosted by AICrowd.com\footnote{\url{http://www.aicrowd.com}}. Inputs from top teams were evaluated 5 times for 100 iterations each in order to achieve a consistent final score. 

\begin{wraptable}{t}{0.3\textwidth}
\vspace*{-0.15in}
\caption{Proportion of algorithmic approaches used, according to a survey conducted with contestant teams.}%
\label{table:algorithmic_approaches}
\vspace*{-0.15in}
{\footnotesize
\begin{tabular}{lc}
\toprule Approach & Proportion \\
\midrule
Bayesian optimization & 34\%  \\
Evolutionary algorithms & 28\% \\
Gradient based & 14\% \\
Meta-heuristics & 7\%  \\
Plackett-Burman design & 3\% \\
Hill climbing & 3\% \\
Other & 10\% \\
\bottomrule
\end{tabular}}
\end{wraptable}

\begin{wrapfigure}{r}{0.45\textwidth}
\centering

\includegraphics[width=0.45\textwidth]{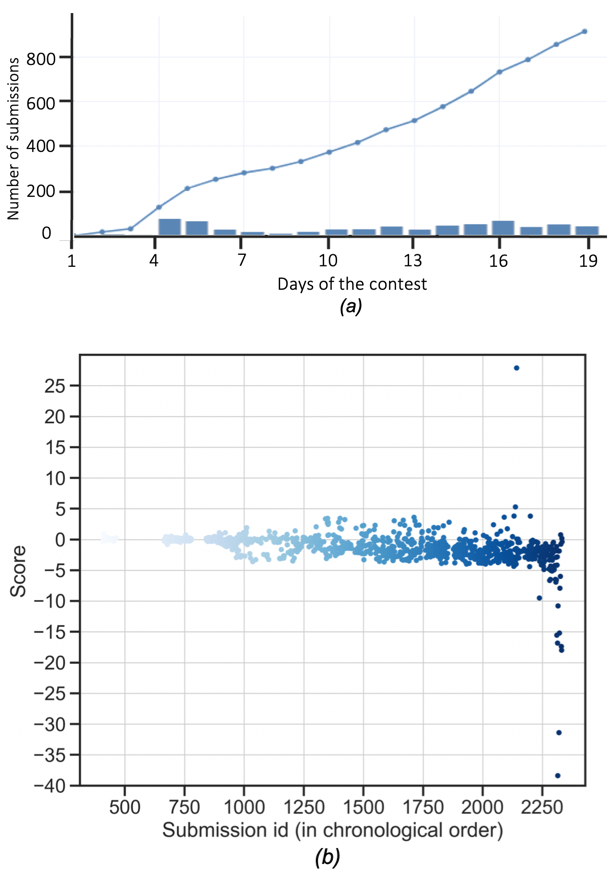}
\caption{Participation history. (a) Number of solutions submitted over time (per day and cumulative). (b) Evolution of scores over time (see equation \cref{eq:zscore_siouxfaux}).}
\label{fig:participation_history}
\end{wrapfigure}

% A score greater than zero represented a solution worse than the \textit{business as usual} scenario (BAU) while a score lower than zero meant an overall improvement of the transportation system over BAU. 
\Cref{fig:participation_history} illustrates the evolution of submissions over time during the competition. Participation developed in two phases. During the first week, contestants became familiar with the BISTRO framework and the Sioux Faux transportation optimization problem. During the second phase, contestants continued to optimize their solutions. 

According to code submissions and a post-contest survey, the solutions that achieved the highest value of the objective function (averaged over 5 replications) followed similar strategies. Typically, they used domain-specific analysis to prune the large input space. For example, by sampling the age distribution within walking distance from a bus route's stops, a reasonable bound on the components of the fare UDI could be assigned. Following these informed factor screening steps, contestants used a variety of algorithmic approaches to more efficiently search the lower dimensional design space. As shown in  \Cref{table:algorithmic_approaches}, the black-box global optimization techniques used during the Contest primarily incorporated variants of Bayesian optimization, genetic/evolutionary algorithms, gradient-based techniques, and meta-heuristics methods.

Most teams managed to improve their scores by three standard deviations better than a random search benchmark (\textit{i.e.}, with scores of approximately -3).  Due to an insidious modeling deficiency, the financial sustainability score component could be optimized towards negative infinity. As such, any contributions from other score components would be relatively inconsequential. Two teams discovered input settings that took advantage of the lack of a lower bound on the financial sustainability score component and were thus able to reach extremely low scores of -30 or -40. This experience underlines the importance of developing careful theory supplemented by judicious testing when designing objective functions.

\paragraph{Post-Contest and New KPIs follow-on studies and results} After the contest, two follow-on studies were conducted to interpret the solutions from the top algorithms in the context of improvements to the objective function. An initial set of improvements, hereby referred to in this text as the ``Post-Contest'' objective function study, simply addressed the unbounded financial sustainability score component as well as other minor problems discovered during the competition.  The ``New KPIs'' objective refers to an expanded set of KPIs, summarized in \cref{table:kpis}. Two of the best-performing algorithms from the Contest---namely, Bayesian Optimization using \textit{tree-based Parzen estimators} (TPE) \cite{Bergstra2011} and \textit{Genetic Algorithms} (GA) \cite{goldberg1988genetic}---were adapted and re-implemented to run BISTRO on the Sioux Faux scenario with both of the updated objective functions. As a baseline algorithmic benchmark, \textit{random search} (RS) was performed for 800 trials using both objective functions. The GA assessment on both the ``Post-Contest'' and ``New KPIs'' objectives utilized five parallel evolutionary trajectories, each drawing a random sample of seeds from a larger gene pool. Both TPE and GA were run over identical design spaces for 1,400 trials on the ``Post-Contest'' and ``New KPIs'' studies\footnote{Partial convergence criteria of 40 iterations were used during initial search, as this was determined to be sufficient for establishing a trajectory consistent with a fully relaxed state.}.

\begin{center}
\begin{figure}[t!]
\centering
 \includegraphics[width=\textwidth]{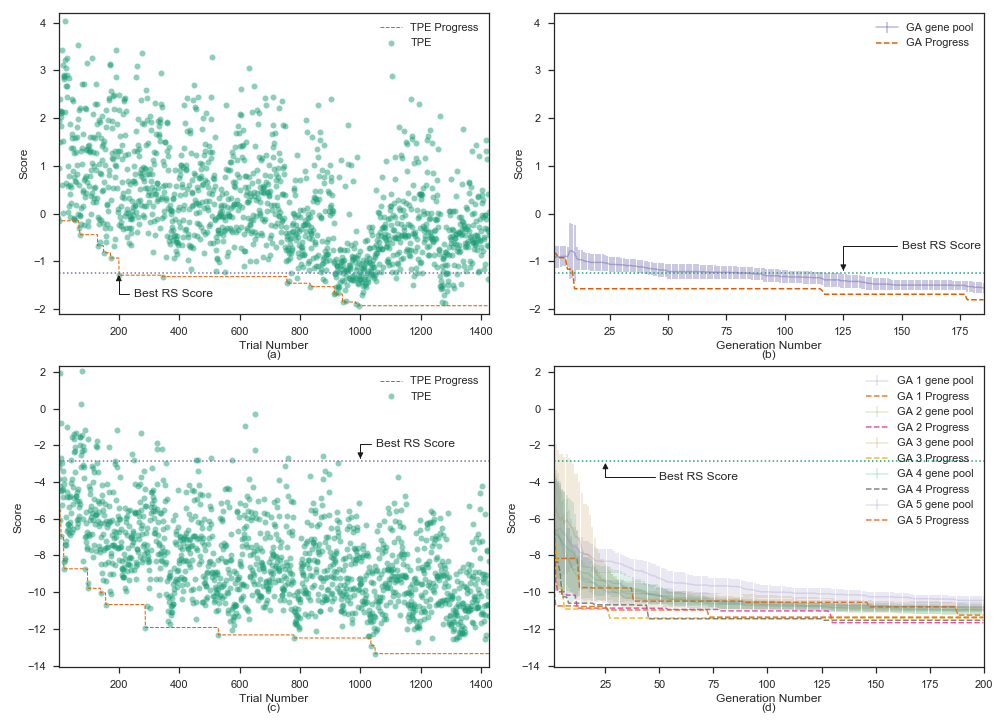}
\caption{Optimization of the Sioux Faux 15k scenario with TPE (left) and GA (right) using ``Post-Contest'' (top) and ``New KPIs'' (bottom) objective function settings. The dashed line(s) across the bottom of each denotes the best (lowest) score achieved by an algorithm within the first $N$ trials. Individual trial scores (at 40 iterations) are shown for TPE plots, whereas one standard deviation ranges of current gene pools are displayed in the GA plots. For the ``Post-Contest'' objective, the TPE and GA algorithms surpass the best score from 800 RS trials of 40 iterations (-1.24) within 200 and 10 trials, respectively. For the ``New KPIs'' objective, both algorithms significantly outperform the best result (-2.84) of an 800 trial, 40 iteration RS almost immediately.}
\label{fig:algo_results}
\end{figure}
\end{center}

% \begin{figure}[ht!]
% \centering
%     \subfigure[ ]{
%         \includegraphics[width=.98\textwidth]{figures/GA_scores.png}
%     }
%     \subfigure[ ]{
%         \includegraphics[width=.98\textwidth]{figures/TPE_scores.png}
%     }
% \caption{Weighted KPI scores for (a) GA and (b) TPE.}
% \label{fig:scores}
% \end{figure}

% The TPE solution runs its buses with medium-to-high fares and medium-to-high incentives across all lines with diverse frequency change choices throughout the day. Conversely, the GA solution offers free transit (for ``Post-Contest'') and mostly free transit (for ``New KPIs'', only charging the 16-65 age group \$2-\$3 on a few bus lines) and does not specify any incentive strategies.

\begin{wrapfigure}{R}{.5\textwidth}
\centering
\vspace*{-0.15in}
    \includegraphics[width=.5\textwidth]{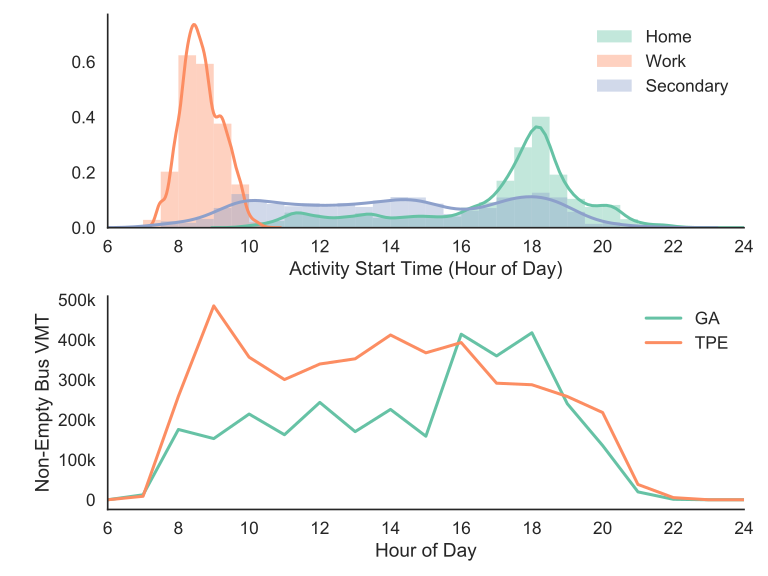}
\caption{Example of output analysis for the ``Post-Contest'' case study. The upper plot shows the various activity start times of agents by activity type. The lower plot shows non-empty bus VMT for two competing algorithms.}
\label{fig:viz_examples}
\end{wrapfigure}
 
Inputs corresponding to the trial yielding the top score for each algorithm were then simulated for 100 iterations with five replicates per trial. \cref{fig:algo_results} demonstrates that both GA and TPE produce input configurations that are superior to RS. We found that both GA and TPE achieve optimal solutions that reflect sensible yet distinct transportation system management strategies. For example, \cref{fig:viz_examples} illustrates the relationship between activity start times (top subfigure) and bus utilization (bottom subfigure) as computed using outputs for the highest scoring solution for GA and TPE algorithms when using the ``Post-Contest'' objective function. This figure suggests that differences in optimal TPE and GA solutions for the ``Post-Contest'' objective arise from distinct transit usage patterns. Note that ``Work'' activity start times occur in the early morning (between 7:00 and 10:00 AM) and correspond to the highest period of utilization for buses under the optimal solution found for the TPE algorithm. In contrast, the GA solution ensures that buses are available during the evening peak commute time; that is, when agents travel back home from work and/or engage in secondary activities. 

\begin{figure}[t]
\centering
 \includegraphics[scale=0.3]{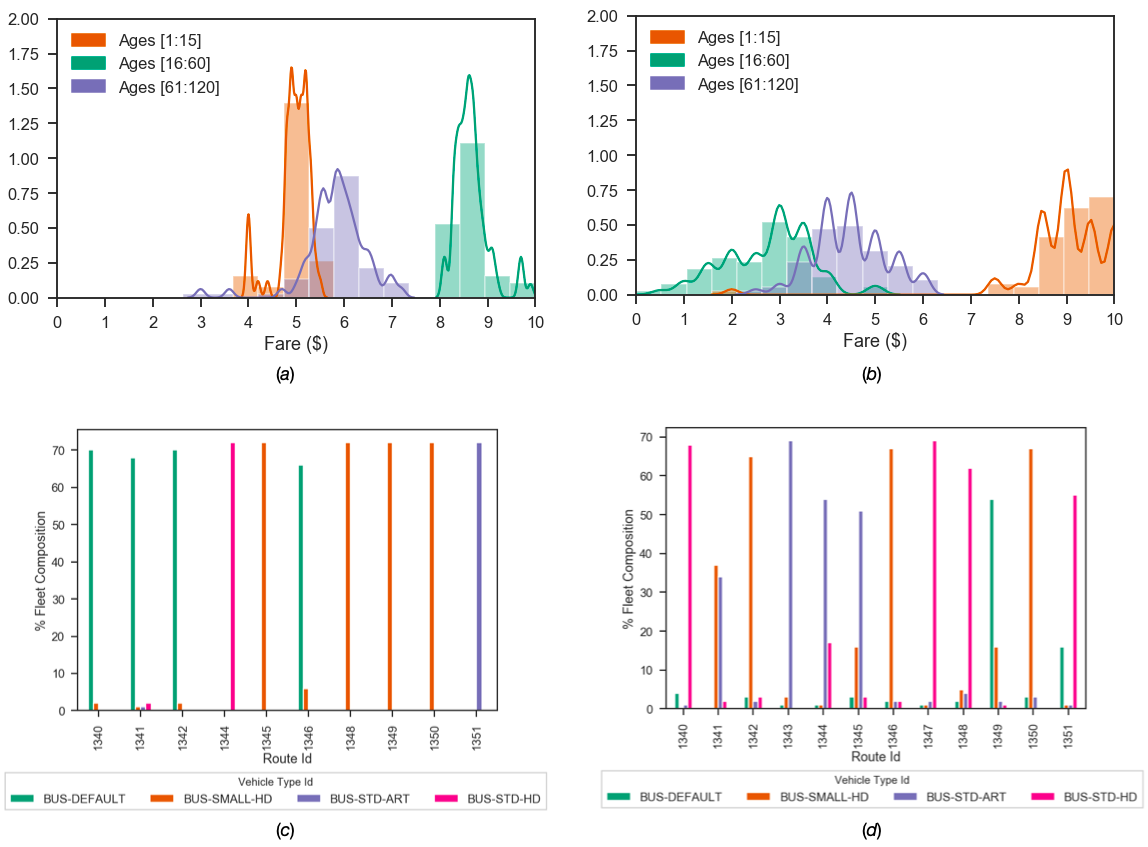}
\caption{Distributions of bus fare by age (top) and vehicle fleet mix by route (bottom) for inputs representing the best fifth percentile scores among trials run for Sioux Faux 15k scenario using the TPE algorithm, shown for ``Post-Contest'' (left) and ``New KPIs'' (right) objective functions.}
\label{fig:algo_interpretability}
\end{figure}

In the case of TPE, we found that the algorithm produced solutions that corresponded to sensible real-world policies.  \cref{fig:algo_interpretability} presents visualizations of input distributions for the top fifth percentile of TPE trials (corresponding to the top 70 of 1400 evaluated solutions by score, as plotted on \cref{fig:algo_results}). This figure illustrates that the values of components of $\hat{d}$ for the best performing (lowest scoring) solutions using TPE occupy a narrow band in the design space. For example, in \cref{fig:algo_interpretability}(a), the highest scoring TPE input value sets evaluated in BISTRO under the ``Post-Contest'' objective suggest charging more expensive bus fares ($\$8-\$10$) for adult citizens (16-60) than for youth (1-15) ($\$4-\$6$) and elderly (60-120) ($\$5-\$7$). The low variance of the components of $\hat{d}$ for these trial points is indicative of both objective function sensitivity to UDI definitions as well as robust algorithm convergence to a (locally) minimal score value. The corresponding bus types on a given route suggested by these solutions are well-resolved and tend towards smaller vehicle models. In contrast, for near-optimal inputs evaluated using the ``New KPIs'' objective function, fares assigned to youth ($\$4-\$10$) are, on average, higher than those assigned to adults ($\$0-\$4$) and seniors ($\$5-\$10$). The corresponding bus types by route are also more diverse among optimal solutions, indicating that the objective is less sensitive to the \texttt{VehicleFleetMix} input when evaluated using the ``New KPI'' objective function. Using the ``New KPIs'' objective, GA (not shown) also finds a tight distribution of fares for top-performing solutions but contrarily finds diversity in its \texttt{VehicleFleetMix} solutions.

\section{Conclusion}
\label{sec:conclusion}
This article has presented the design, software architecture, and preliminary evaluation of BISTRO: a general-purpose transportation policy decision support tool and scenario-based optimization framework supported by empirically-driven agent-based models. When combined with sensible guidance from experienced planning professionals, BISTRO can be used to identify more holistic, empirically-driven approaches to urban transportation planning and management. In addition to overall system purpose, design, and software architecture, this work provides a concrete example of the process that BISTRO supports as implemented in the context of a scenario-based policy optimization ``contest.'' While many participants had little or no prior expertise in the transportation science and policy analysis methods typically used in urban planning practice, over a dozen teams developed algorithms that found inputs, which, when evaluated in the simulator, achieved scores that surpassed both random search as well as human judgment.  The mixed results of the competition led us to conclude that the optimization-based search techniques enabled by BISTRO should support an iterative approach that involves applying optimization algorithms to refinements of KPI specifications in order to better align objective functions with system goals.

 Research conducted using BISTRO strives to meet the highest standard of reproducibility in computational experiments \cite{peng2011reproducible,morin2012shining,sandve2013ten} as well as fact-based policymaking \cite{hastings2019} by making all data, models, and algorithms freely available and open source\footnote{All code and data used and referred in the article is available at \url{http://bistro.its.berkeley.edu}}. One finding of post-contest reproducibility efforts was that different classes of algorithms appeared to converge to solutions that emphasized distinct policy strategies.

Our experience from this pilot study demonstrates that we have implemented a compelling platform to study human-in-the-loop design of expensive simulation-based optimization algorithms. This conclusion suggests that, in addition to its utility as a decision support system, BISTRO could serve as an exemplary testbed for multiple emerging streams of research (\textit{e.g.}, freeze-thaw, multi-objective, multi-task, and multi-fidelity optimization) in SMBO and associated meta-model-based optimization methods. Should BISTRO be widely adopted as part of the urban planning toolkit, innovative algorithms and new theory developed as part of inquiry in these sub-domains will have the added benefit of directly serving a humanitarian purpose.

\begin{acks}
The authors wish to thank the Uber employees who participated in the contest and provided valuable feedback.
\end{acks}

\bibliography{main.bib}
\bibliographystyle{unsrt}

\end{document}